\pdfoutput=1 
\documentclass[a4paper,12pt]{article}
\usepackage{preamble}

\usepackage[sorting=none,style=numeric-comp,maxbibnames=6]{biblatex}
\renewcommand{\baselinestretch}{1.2}
\makeatletter
\def\blfootnote{\gdef\@thefnmark{}\@footnotetext}
\makeatother

\begin{document}

\thispagestyle{empty}
\vspace{13mm}  
\begin{center}
{\huge  The Bethe Ansatz for the superconformal}
\\[5mm]
{\huge index with unequal angular momenta}
\\[13mm]
{\large Ofer Aharony$^{1,2}$, Ohad Mamroud$^{1,3,4}$, Shimon Nowik$^1$, Meir Weissman$^1$}
 
\bigskip
{\it
$^1$ Department of Particle Physics and Astrophysics, \\[.0em]
Weizmann Institute of Science, Rehovot 7610001, Israel \\[.2em]
$^2$ School of Natural Sciences, Institute for Advanced Study, Princeton 08540, NJ, USA \\[.2em]
$^3$ SISSA, Via Bonomea 265, 34136 Trieste, Italy \\[.2em]
$^4$ INFN, Sezione di Trieste, Via Valerio 2, 34127 Trieste, Italy 
}

\bigskip
\bigskip

{\parbox{16cm}{\hspace{5mm}
A few years ago it was shown that the superconformal index of the ${\cal N}=4$ supersymmetric $SU(N)$ Yang-Mills theory in the large $N$ limit matches with the entropy of $1/16$-supersymmetric black holes in type IIB string theory on $AdS_5\times S^5$. In some cases, an even more detailed match between the two sides is possible. When the two angular momentum chemical potentials in the index are equal, the superconformal index can be written as a discrete sum of Bethe ansatz solutions, and it was shown that specific terms in this sum are in a one-to-one correspondence to stable black hole solutions, and that the matching can be extended to non-perturbative contributions from wrapped D3-branes. A Bethe ansatz approach to computing the superconformal index exists also when the ratio of the angular momentum chemical potentials is any rational number, but in those cases it involves a sum over a very large number of terms (growing exponentially with $N$). Benini et al showed that a specific one of these terms matches with the black hole, but the role of the other terms is not clear. In this paper we analyze some of the additional contributions to the index in the Bethe ansatz approach, and we find that their matching to the gravity side is much more complicated than in the case of equal chemical potentials. In particular, we find some contributions that are larger than the one which was found to match the black holes, so that they must cancel with other large contributions. We give some evidence that cancellations of this type are possible, but we leave a full understanding of how they work to the future.
}}
\end{center}

\newpage
\pagenumbering{arabic}
\setcounter{page}{1}
\setcounter{footnote}{0}
\renewcommand{\thefootnote}{\arabic{footnote}}

{\renewcommand{\baselinestretch}{.88} \parskip=0pt
\setcounter{tocdepth}{2}
\tableofcontents}


\section{Introduction}
\label{sec:intro}

Recent years have seen significant advances in precision holography, where semi-classical bulk information was precisely reproduced by a computation in the dual conformal field theory (CFT), mostly in a supersymmetric setting. The first breakthroughs were in counting the microstates of supersymmetric black holes by analyzing various supersymmetric indices, thus reproducing their entropy \cite{Benini:2015eyy, Cabo-Bizet:2018ehj, Choi:2018hmj, Benini:2018ywd}. Further improvements followed, generalizing these results to many different theories and computing more refined data on both sides of the correspondence \cite{Aharony:2021zkr,Honda:2019cio, ArabiArdehali:2019tdm, Kim:2019yrz, Cabo-Bizet:2019osg, Amariti:2019mgp, Lezcano:2019pae, Lanir:2019abx, Cabo-Bizet:2019eaf, ArabiArdehali:2019orz, Cabo-Bizet:2020nkr, Murthy:2020rbd, Agarwal:2020zwm, Benini:2020gjh, GonzalezLezcano:2020yeb, Copetti:2020dil, Cabo-Bizet:2020ewf, Amariti:2020jyx, Hosseini:2016tor, Benini:2016hjo, Benini:2016rke, Hosseini:2016cyf, Cabo-Bizet:2017jsl, Azzurli:2017kxo, Hosseini:2017fjo, Benini:2017oxt, Hosseini:2018uzp, Crichigno:2018adf, Hosseini:2018usu, Suh:2018szn, Fluder:2019szh, Gang:2019uay, Kantor:2019lfo, Choi:2019zpz, Choi:2021rxi, Bobev:2019zmz, Nian:2019pxj, Benini:2019dyp, ArabiArdehali:2021nsx,David:2021qaa, David:2020ems, Amariti:2022nvn, Hosseini:2021mnn, Hong:2021bzg, Goldstein:2020yvj, Jejjala:2021hlt, Jejjala:2022lrm, Mamroud:2022msu,BenettiGenolini:2023rkq,Choi:2023tiq,Bobev:2023dwx,Cabo-Bizet:2023ejm,Chen:2023lzq}.

One particularly convenient set-up to study was that of the four dimensional $\cN = 4$ $SU(N)$ supersymmetric Yang-Mills theory, where the superconformal index of the theory can be matched to the dual gravitational partition function. While the index is usually expressed via an integral formula, the \emph{Bethe Ansatz approach} \cite{Benini:2018mlo,Closset:2017bse} allows us in some cases to localize the integral and to transform it into a discrete sum (see Section~\ref{sec:Review of BA} for details and caveats), schematically
\begin{equation}
\label{eq:Schematic BA sum}
    \cI = \sum_{u \in BA} \cI_u \,,
\end{equation}
where the $u$'s can be thought of as specific configurations of the complexified holonomies of the $SU(N)$ gauge field which solve some set of transcendental equations (and the sum is over the set of solutions to these equations). In \cite{Benini:2018ywd, Aharony:2021zkr} a particular family of terms in the sum was analyzed, and at large $N$ each of them contributed to the index at order $e^{\# N^2}$. These turned out to precisely match the contributions of various different gravitational saddle points in the bulk dual, in a one-to-one fashion (and including some order-$N$ corrections coming from wrapped D-branes).

Unfortunately, the Bethe Ansatz approach can only compute the index for some particular choices of the chemical potentials appearing in the superconformal index, those where the two angular chemical potentials of the theory $\tau$ and $\sigma$ have a rational ratio, $\tau = a \omega$ and $\sigma = b\omega$ for $a,b\in\bN$. Each contribution to the index, $\cI_u$, is then described as a sum over $(ab)^{N-1}$ terms, where one shifts the holonomies of the solution $u$ in some prescribed way.

The detailed analysis of the matching to gravity in \cite{Aharony:2021zkr} concerned an even simpler case, that in which $\sigma=\tau$, for which each $\cI_u$ is given by exactly one term. In the more general case, \cite{Benini:2020gjh} analyzed one of the $(ab)^{N-1}$ terms in the sum (for a particular solution to the Bethe Ansatz equations) and showed that it reproduces the gravitational action of a specific black hole solution. Another term was analyzed (for more general solutions of the Bethe Ansatz equations) in \cite{Colombo:2021kbb} and was also found to reproduce the gravity answers. A natural question is then what happens to the other terms at large $N$? Do they give negligible contributions, do they cancel amongst themselves, or do they match some other gravitational background?

In this work we analyze some specific additional terms in the sum at large $N$, as depicted in Figure~\ref{fig:different Ms}. As opposed to the cases analyzed in \cite{Benini:2020gjh,Colombo:2021kbb}, here we consider configurations of the holonomies which do not correspond at large $N$ to uniform distributions on cycles of the complex torus with modular parameter $ab\omega$. We find that not only do they contribute at order $e^{\# N^2}$, but they do so in a way that is different from any gravitational background familiar to us. In some cases their contribution is even larger than that of the terms analyzed in \cite{Benini:2020gjh,Colombo:2021kbb}. However, we did not analyze all the different $(ab)^{N-1}$ terms, and it is possible that they cancel amongst themselves, or with other terms in \eqref{eq:Schematic BA sum}.

The paper is organized as follows: in Section~\ref{sec:Review of BA} we review the Bethe-Ansatz approach. In Section~\ref{sec:Reduced BA} we generalize a proof of \cite{Benini:2021ano} to show that only solutions to the reduced Bethe Ansatz equations contribute to the index, also when the angular chemical potentials are different. In Section~\ref{sec:Large N expansion} we compute the large $N$ expansion of the aforementioned configurations. In Section~\ref{sec:su2} we study the case $N=2$; this case is very far from the large $N$ limit, but in this case we can analyze all the $(ab)$ different terms contributing to $\cI_u$, and we show that cancellations between different terms are plausible. In Section~\ref{sec:shifts from gravity} we analyze the known gravitational saddles, and show that they cannot all be explained by a large $N$ limit of one of the known configurations. An appendix contains some details on the special functions which appear in the computations.

\section{Review of the Bethe Ansatz method}
\label{sec:Review of BA}
The superconformal index can be defined for any four dimensional $\cN = 1$ superconformal field theory. One starts by picking a Poincar\'e supercharge $\cQ$ and its superconformal conjugate $\cQ^\dagger$. The superconformal index then counts the difference between bosonic and fermionic supersymmetric operators that are annihilated by this supercharge. One can add fugacities for charges that commute with the relevant supercharge. The superconformal index can then be written as a trace over the entire Hilbert space of states on $S^3$. For the case of $\cN = 4$ super Yang-Mills, the index counts $1/16$-BPS operators and is calculated via\footnote{Often the powers of the fugacities are written as $p^{J_1+\frac12 r} \, q^{J_2+\frac12 r}$. Compared to this convention, we have swallowed a power of $(pq)^{1/3}$ into $y_1$ and $y_2$, such that the index is a single-valued function. The relation of our variables to those of \cite{Kinney:2005ej} is $p = t^3y \big|_\text{there}$, $q = t^3/y \big|_\text{there}$, $y_1 = t^2v \big|_\text{there}$, $y_2 = t^2 w/v \big|_\text{there}$.}
\begin{equation}
    \cI(y_{1,2},p,q) = \Tr\left[(-1)^F e^{-\{\cQ,\cQ^\dagger\}} p^{J_1 + \frac{1}{2}R_3} q^{J_2 + \frac{1}{2}R_3} y_1^{\frac{1}{2}(R_1 - R_3)} y_2^{\frac{1}{2}(R_2 - R_3)} \right] \;,
\end{equation}
where $R_{1,2,3}$ are in the Cartan of the $SU(4)_R$ R-symmetry, with the three complex scalars of the theory having charge 2 under one of them and zero under the other two, in a symmetric way. $J_{1,2}$ are the half-integer angular momenta quantum numbers of local operators, each rotating an $\bR^2 \subset \bR^4$. $F$ is the fermion number. $\cQ$ is chosen to be the complex supercharge associated with the R-symmetry generator $r = \frac{1}{3}(R_1 + R_2 + R_3)$. 

We find it convenient to write the index in terms of chemical potentials instead of fugacities, which are denoted by
\begin{equation}
    p = e^{2\pi i \sigma} \,,\qquad q = e^{2\pi i \tau}\,,\qquad y_{1,2} = e^{2\pi i \Delta_{1,2}}\;.
\end{equation}
The index is well defined for $|p|,|q|<1$, i.e. for $\im(\sigma),\im(\tau) > 0$.

It has long been known that the index has an integral form \cite{Kinney:2005ej,Romelsberger:2005eg}, in which it is expressed in terms of a contour integral 
over the Cartan subalgebra of the gauge group, in our case $SU\left(N\right)$,
\begin{equation}
\cI\left(\Delta_{1,2},\tau,\sigma\right) = \kappa_{N}\left(\Delta_{1,2};\tau,\sigma\right) 
 \oint_{\mathbb{T}^{N-1}} \left(\prod_{i=1}^{N-1}du_{i} \right) \cZ\left(\{u_i\},\Delta_{1,2};\tau,\sigma\right)\;,
\label{eq: index as integral-1}
\end{equation}
where
\begin{align}
    \label{eq:def cZ}
    \cZ\left(\left\{ u_{i}\right\} , \Delta_{1,2} ;\tau,\sigma\right) &= \prod_{i\neq j}^{N}\frac{\tilde{\Gamma}\left(u_{ij}+\Delta_{1};\tau,\sigma\right)\tilde{\Gamma}\left(u_{ij}+\Delta_{2};\tau,\sigma\right)}{\tilde{\Gamma}\left(u_{ij};\tau,\sigma\right)\tilde{\Gamma}\left(u_{ij}+\Delta_{1}+\Delta_{2};\tau,\sigma\right)} \;, \\
    \label{eq:def k_N}
    \kappa_{N}\left(\Delta_{1,2} ;\tau,\sigma\right) &= \frac{1}{N!}\left(\frac{\left(p;p\right)_{\infty}\left(q;q\right)_{\infty}\tilde{\Gamma}\left(\Delta_{1};\tau,\sigma\right)\tilde{\Gamma}\left(\Delta_{2};\tau,\sigma\right)}{\tilde{\Gamma}\left(\Delta_{1}+\Delta_{2};\tau,\sigma\right)}\right)^{N-1} \;.
\end{align}
Here
$u_{ij}=u_{i}-u_{j}$, $\left(z;q\right)_{\infty}$ is the $q$-Pochhammer
symbol, and $\tilde{\Gamma}$ is the elliptic gamma
function, both defined in Appendix \ref{app:Special functions}.
The variables $u_{i}$ parameterize the torus $\mathbb{T}^{N-1}$
on which we integrate. $u_{N}$ is defined such that the $e^{2\pi iu}$'s can be thought of as the eigenvalues of a special unitary matrix, i.e. via the constraint
\begin{equation}
\label{eq: SU(N) constraint}
\sum_{i=1}^{N}u_{i} = 0\;.
\end{equation}

There are various methods for calculating the contour integral \eqref{eq: index as integral-1}. In this paper we focus on the Bethe Ansatz (BA) approach \cite{Benini:2018mlo,Closset:2017bse}, which
is valid at finite $N$, and then expand at large $N$. This formula manipulates the integrand and the contour such that the integral can be written as a sum of residues.

The BA formula applies when the chemical potentials $\tau,\sigma$ have a rational ratio, i.e. $\tau = a \omega$, $\sigma = b\omega$, for some $a,b \in \bN$ such that $\gcd(a,b)=1$, $\im (\omega) > 0$. We also define the fugacity 
\begin{equation}
h = e^{2\pi i\omega}\,, \qquad p = h^b\,,\qquad q = h^a \;.
\end{equation}

\paragraph{Derivation of the BA formula}

Following \cite{Benini:2018mlo}, we can manipulate the contour integral \eqref{eq: index as integral-1} by noting that the integrand, $\cZ$, is a quasi-elliptic function\footnote{To prove this, use (3.38) of \cite{Benini:2018mlo} for every term in $\cZ$.},
\begin{equation} \label{qzz}
    (-1)^{N-1}Q_j\left(\{u_i\}, \Delta_{1,2};\omega \right) \cZ\left(\{u_i\}, \Delta_{1,2};a\omega,b\omega \right) = \cZ\left(\left\{ u_{i} - \delta_{ij} ab\omega \right\} , \Delta_{1,2} ; a\omega, b\omega\right) \;,
\end{equation}
where for $i=1,\cdots,N$ we define
\begin{equation}
\label{eq: quasi periodicity}
    Q_i\left(\{u_j\},\Delta_{1,2};\omega\right) \equiv e^{6\pi i \sum_{j=1}^N u_{ij}} \prod_{j=1}^N \frac{\theta_0(\Delta_1+u_{ji};\omega) \theta_0(\Delta_2+u_{ji};\omega) \theta_0(-\Delta_1-\Delta_2+u_{ji};\omega)}{\theta_0(\Delta_1-u_{ji};\omega) \theta_0(\Delta_2-u_{ji};\omega) \theta_0(-\Delta_1-\Delta_2-u_{ji};\omega)} \;.
\end{equation}
Note that acting with this operator shifts one of the $u_i$'s and therefore breaks the $SU(N)$ constraint \eqref{eq: SU(N) constraint}. However, acting with 
\begin{equation}
    \hat Q_i = \frac{Q_i}{Q_N} \;
\end{equation}
preserves the constraint. We note that the $Q_i$'s are elliptic functions in each of the $u_i$'s separately, i.e. they are periodic under $u_i \sim u_i + 1 \sim u_i+\omega$.

We can now use \eqref{qzz} to modify the contour of
integration given in \eqref{eq: index as integral-1},
\begin{equation}
\begin{aligned}
\cI\left(\Delta_{1,2};a\omega,b\omega\right) &= \kappa_{N} \oint_{\mathbb{T}^{N-1}} \left(\prod_{i=1}^{N-1}du_{i}\right) \prod_{i=1}^{N-1}\frac{1-\hat{Q}_{i}\left(\left\{u_{j}\right\},\Delta_{1,2};\omega\right)}{1-\hat{Q}_{i}\left(\left\{u_{j}\right\},\Delta_{1,2};\omega\right)} \cZ\left(\left\{u_{i}\right\},\Delta_{1,2};a\omega,b\omega\right) \\
&= \kappa_{N} \oint_{\cC} \left(\prod_{i=1}^{N-1}du_{i}\right) \frac{\cZ\left(\left\{u_{i}\right\},\Delta_{1,2};a\omega,b\omega\right)}{\prod_{i=1}^{N-1}\left(1-\hat{Q}_{i}\left(\left\{u_{j}\right\},\Delta_{1,2};\omega\right)\right)}
\end{aligned}
\end{equation}
where the new integration contour, $\cC$, is the contour encircling the annulus 
\begin{equation}
\mathcal{A}=\left\{ u_{i}\left|1<\left|e^{2\pi iu_{i}}\right|<\left|h\right|^{-ab},i=1,\cdots,N-1\right.\right\}  \;.
\end{equation}

Applying the residue theorem\footnote{As shown in \cite{Benini:2018mlo}, no other poles in the integrand will contribute, since they are
either canceled by poles of the denominator of a high enough
degree, or they are outside the new contour of integration.}, 
\begin{equation}
\mathcal{I}\left(\Delta_{1,2} ;a\omega,b\omega\right) = \kappa_{N} \sum_{\left\{u_{i}\right\} \in\text{BAEs}} \sum_{\left\{ m_{i}\right\}=1}^{ab} \cZ\left(\left\{u_{i}-m_{i}\omega\right\},\Delta_{1,2};a\omega,b\omega\right) \cdot H^{-1}\left(\left\{u_{i}\right\}, \Delta_{1,2};\omega\right),
\label{eq: BA formulation sum-1}
\end{equation}
where the sum is over all solutions to the BA equations\footnote{This is a bit of a simplification, since we assumed that all solutions are discrete and not part of a continuum. That is known not to be the case for $N \geq 3$. The contribution of continuous families of solutions will be discussed in \cite{Aharony:2024}. It does not seem to qualitatively change the behavior discussed in this paper.}
\begin{equation}
\label{eq:BA equations}
\hat Q_i\left(\left\{u_{j}\right\},\Delta_{1,2};\omega\right) = 1 \;,
\end{equation}
and $H^{-1}$ is the Jacobian coming from the change of variables from the $u_j$ to the $\hat{Q}_{i}$'s when we apply the residue theorem to the integral.
By $\sum_{\left\{ m_{i}\right\} =1}^{ab}$ we mean $\sum_{m_1=1}^{ab} \cdots \sum_{m_{N-1}=1}^{ab}$, with $m_N = -\sum_{k=1}^{N-1} m_k$, coming from all the different shifts of the $u_i$'s by $\omega$ that are within $\cA$, as the BA equations \eqref{eq:BA equations} are periodic under these shifts.
In the Bethe Ansatz approach we pick up the poles of the integrand in the region $0 > \im\left(u_{i}\right) \geq \im\left(-ab\omega\right)$, while every solution to \eqref{eq:BA equations} can be chosen to be inside the torus with parameter $\omega$ (namely, such that for all $i=1,\cdots,N-1$ we have $\im(\omega)>\im(u_{i})\ge0$). As a consequence, each such solution corresponds to $(ab)^{N-1}$ poles of the integrand, where we shift each $u_{i}$ by $(-m_{i}\omega)$ for $i=1,\cdots,N-1$, $m_{i}=1,\cdots,ab$, with a compensating shift for $u_{N}$.

For $\sigma=\tau$, $a=b=1$ and the sum over $\{m_i\}$ is trivial. However, for any other case $ab>1$, and
the number of terms in the sum, $(ab)^{N-1}$, is exponential in $N$. We stress that the dependence on $a$, $b$ and the $m_i$
is present only in $\mathcal{Z}$, and not in $\kappa_N$ or $H^{-1}$, so this complication will
only affect the ability to calculate $\mathcal{Z}$.

\paragraph{Hong-Liu solutions}
While the full set of solutions to the BAEs for $N>2$ is unknown,
a specific set of solutions, named the Hong-Liu (HL) solutions, was
identified \cite{Hong:2018viz}. They correspond to symmetric configurations of the $u$'s on the $(1,\omega)$ torus, and are denoted by three integers, $\left\{ m,n,r\right\} $,
such that $N=m\cdot n$, and $r=0,\cdots,n-1$. Their explicit form is 
\begin{equation}
\label{eq:HL solutions}
u_{j} \equiv u_{\hat{\jmath
}\hat{k}} = \bar{u} + \frac{\hat{\jmath}}{m} + \frac{\hat{k}}{n} \left(\omega+\frac{r}{m}\right) 
\end{equation}
such that $\hat{\jmath}=0,\cdots,m-1$ and $\hat{k}=0,\cdots,n-1$, and $\bar{u}$
is a constant chosen to satisfy the $SU(N)$ constraint \eqref{eq: SU(N) constraint}. The $\left\{ 1,N,0\right\}$
solution is sometimes named the ``basic solution'', and it is given by
\begin{equation}
\label{eq: basic solution}
u_{j} = \bar{u} + \frac{j\omega}{N}\;.
\end{equation}
Different HL solutions and different series of $\left\{ m_{j}\right\}$,
have a different degree of dominance in the sum \eqref{eq: BA formulation sum-1}, and most of these
contributions have not been calculated yet.

\section{The reduced Bethe Ansatz suffices}
\label{sec:Reduced BA}


As discussed in \cite{Benini:2021ano}, in addition to the shifts of $u_i$ by $(-m_i \omega)$ discussed above, every BA solution is related to $N^2$ additional solutions, by shifting all $u_{i}\to u_{i}+\frac{\alpha+\beta\omega}{N}$ $(i=1,\cdots,N-1)$, and $u_{N}\to u_{N}+\left(1-N\right)\frac{\alpha+\beta\omega}{N}$, where $\alpha,\beta=0,\cdots,N-1$. The differences $u_{ij}$ are affected if and only if either $i=N$ or $j=N$, and the $u_{iN}$ are just shifted by $(\alpha+\beta\omega)$, which leaves the Bethe Ansatz equations invariant\footnote{Note that in some cases the shift may take a solution to itself up to a permutation of the eigenvalues, but this will not affect the arguments below.}. 

If a shift by some $(\frac{\beta}{N}-m_i) \omega$ takes us outside of the integration region, we can shift for that specific shifted solution $u_{i}\to u_{i}-ab\omega$ and $u_{N}\to u_{N}+ab\omega$ to return back into it. As mentioned in the previous section, this leaves the Jacobian invariant and multiplies $\cZ$ by $\hat Q_{i}$. Since we started from a solution to the BAE and since $\hat{Q}_{i}$ has periodicity $\omega$, $\hat{Q}_{i}=1$ in our case and the contribution to the index is invariant under this shift. Thus, in computing the index we can keep the original shifts $u_{i}\to u_{i}+\frac{\alpha+\beta\omega}{N}$ and we do not need to worry about this issue. In fact, by the same arguments, we can consider such shifts with $\beta=0,1,\cdots,abN-1$. The extra shifts we added by integer multiples of $\omega$ do not give new solutions but rather different $m_i$-shifts of the same solutions, so adding these extra shifts just multiplies the result by $ab$, but it will be useful to write the index as a sum over these shifts for the argument below. 

Note that since $\cZ$ only depends on the differences of the $u_i$ and does not depend on the constraint $\sum_{i=1}^N u_i=0$, for computing $\cZ$ we can equally well describe these shifts as acting only on $u_{N}$, namely
\begin{equation}
u_{i}\to\begin{cases}
u_{i} & i\neq N,\\
u_{N} - \alpha - \beta\omega & i=N.
\end{cases}
\end{equation}

Since both $\cZ$ and the Jacobian $H^{-1}$ are invariant under integer shifts of the $u_{ij}$'s, summing over the $\alpha$ shifts will just give a multiplicative factor of $N$. In the case $\tau = \sigma$ the authors of \cite{Benini:2021ano} showed that due to the $\beta$ shifts, solutions for the Bethe ansatz equations $\{ \hat{Q}_i = 1;\ i=1,\cdots,N-1\}$ which are not solutions to the reduced Bethe ansatz equation $\{ Q_i = (-1)^{N-1};\  i=1,\cdots,N \}$ cancel each other, such that the index only gets contributions from solutions to the reduced BAE. We will now argue that the same is true also for $\tau\neq \sigma$.

Picking a solution $\left\{ u_{i}\right\}$ to the BAE, as described above we can sum
over the shifts $\beta=0,\cdots,abN-1$ (in addition to the $m_{j}$ shifts).
The Jacobian $H^{-1}$ is invariant under $\omega\bZ$ shifts of the $u$'s, so we will just need to evaluate the effect of summing over the $\beta \omega$ shifts in $\cZ$. The contribution of the solution (and its shifted relatives) to the index is therefore
\begin{equation}
\cI_{\{u\}} = \frac{N}{ab} \cdot\kappa_N \cdot H^{-1}\left(\{u_i\},\Delta_{1,2};\omega\right)\sum_{\left\{ m_{j}\right\} =1}^{ab} \sum_{\beta=0}^{abN-1} \cZ\left(u_{i}-m_{i}\omega,\,u_{N}+\sum_{i=1}^{N-1}m_{i}\omega-\beta\omega;\Delta_{1,2},a\omega,b\omega\right) \;.
\end{equation}
We can write $\beta=ab\beta_1+\beta_2$ with $\beta_1=0,\cdots,N-1$ and $\beta_2=0,\cdots,ab-1$,
\begin{equation}
\cI_{\{u\}} = \frac{N}{ab} \cdot\kappa_N \cdot H^{-1} \cdot \sum_{\left\{ m_{j}\right\} =1}^{ab} \sum_{\beta_2=0}^{ab-1}
\sum_{\beta_1=0}^{N-1} \cZ\left(u_{i}-m_{i}\omega,\,u_{N}+\sum_{i=1}^{N-1}m_{i}\omega-\beta_2\omega - \beta_1 ab\omega;\Delta_{1,2},a\omega,b\omega\right) \;.
\end{equation}
But since shifting only $u_{N}\to u_{N}-ab\omega$ amounts to multiplying $\cZ$ by $\left(-1\right)^{N-1}Q_{N}$, as in \eqref{qzz}, and since $Q_{N}$ has $\omega$ periodicity in each $u_{i}$ separately, this is the same as
\begin{multline}
\label{eq:contribution of BAE solution with multiplicity}
\cI_{\{u\}} = \frac{N}{ab} \cdot\kappa_N \cdot H^{-1}
\cdot \sum_{\left\{ m_{j}\right\} =1}^{ab} \sum_{\beta_2=0}^{ab-1}
\sum_{\beta_1=0}^{N-1} ((-1)^{N-1} Q_N(u_i))^{\beta_1} \\ \times \cZ\left(u_{i}-m_{i}\omega,\,u_{N}+\sum_{i=1}^{N-1}m_{i}\omega-\beta_2\omega;\Delta_{1,2},a\omega,b\omega\right) \;.
\end{multline}
Since $\hat Q_i$ solves the Bethe ansatz equations and $\prod_{i=1}^N Q_i = 1$, $Q_{N}$ is an $N$-th root of unity, and so the sum over $\beta_1$ vanishes unless $Q_{N}=\left(-1\right)^{N-1}$. Since all the $Q_{i}$'s are equal to each other, we see that only solutions to the reduced Bethe Ansatz equation 
\begin{equation}
Q_{i} = \left(-1\right)^{N-1}
\end{equation}
contribute to the index.

\section{Large $N$ expansion}
\label{sec:Large N expansion}

Here we will be interested in the large $N$ contribution of the Hong-Liu solutions to the $SU(N)$ index, and in particular the contribution of the basic solution \eqref{eq: basic solution}. We will concentrate on the leading order in $N$ -- terms that are exponential in $N^2$, whose log is $O(N^2)$. As explained previously, there are $(ab)^{N-1}$ different terms in this contribution, coming from the possible shifts by $\{m_i\omega\}_{i=1}^{N-1}$ of the basic solution. Therefore, an exponential in $N^2$ dependence must come from individual shifts, and not from the summation over them. Each choice for the $m$'s gives
\begin{equation}
    \log \left(\cI_{\{u\}}(\{m_i\})\right) = \log\left(\kappa_N\right) + \log \left(H^{-1}\right) + \log \left(\cZ\left(\{u_{i}-m_{i}\omega\},\,u_{N} + \sum_{i=1}^{N-1}m_{i}\omega;\Delta,a\omega,b\omega\right)\right) \;.
\end{equation}
The first term does not contribute at order $O(N^2)$, as is obvious from \eqref{eq:def k_N}. The second term was shown to be $O(1)$ in
\cite{Benini:2018ywd}, and was exactly computed in \cite{Mamroud:2022msu} for the HL solutions with $m=0$. So we are left with evaluating the $N$ dependence of the third term, $\log (\cZ)$. 

Evaluating the last term depends significantly on the choice of $m$'s. In \cite{Benini:2020gjh}, the authors evaluated it for the choice $m_j = j \mod ab$, and showed that\footnote{In some regimes of the chemical potentials $\Delta$ we need to replace $\left[\Delta\right]_{\omega}$ by $ \left[\Delta\right]_{\omega}^\prime =  \left[\Delta\right]_{\omega} + 1$.}
\begin{equation} \label{Mequalsone}
    \log \left[\cI_{\{1,N,0\}}\left(\left\{m_j = j\right\}\right)\right] = -\pi i N^2 \frac{\left[\Delta_1\right]_\omega \left[\Delta_2\right]_\omega \left[\Delta_3\right]_\omega }{a b \omega^2} + O\left(N\log (N)\right)\;,
\end{equation}
which agrees with the on-shell action of dual black holes, where $[\Delta_3]_\omega = \tau + \sigma - [\Delta_1]_\omega - [\Delta_2]_\omega - 1$ and the function $[\Delta]_\omega$ simply shifts $\Delta$ to a strip, such that it satisfies
\begin{equation}
    \label{eq:def bracket}
    [\Delta]_\omega = \Delta \mod 1 \,, \qquad \text{such that} \qquad -\im\left(\frac{1}{\omega}\right) > \im\left(\frac{[\Delta]_\omega}{\omega}\right) > 0 \,.
\end{equation}

\begin{figure}[t]
\centering
\begin{subfigure}[]{0.3 \textwidth}
\includegraphics[width=\textwidth]{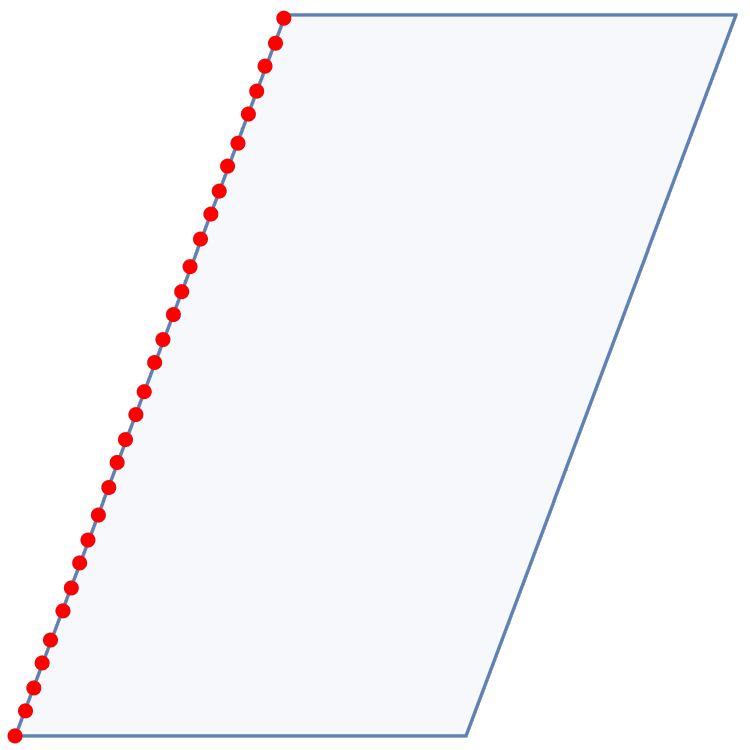}
\caption{$M = 1$}
\end{subfigure}
\hfill
\begin{subfigure}[]{0.3 \textwidth}
\includegraphics[width=\textwidth]{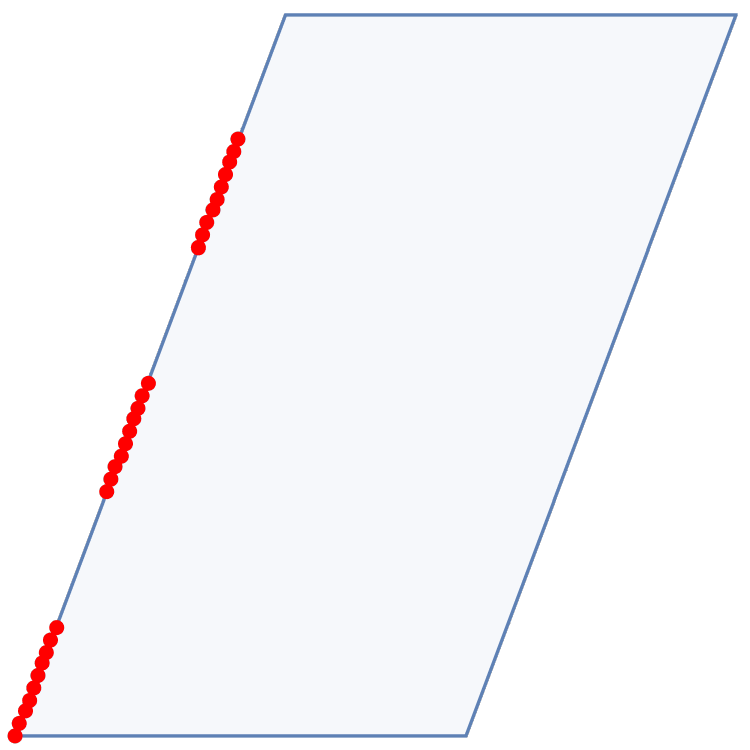}
\caption{$M = 2$}
\end{subfigure}
\hfill
\begin{subfigure}[]{0.3 \textwidth}
\includegraphics[width=\textwidth]{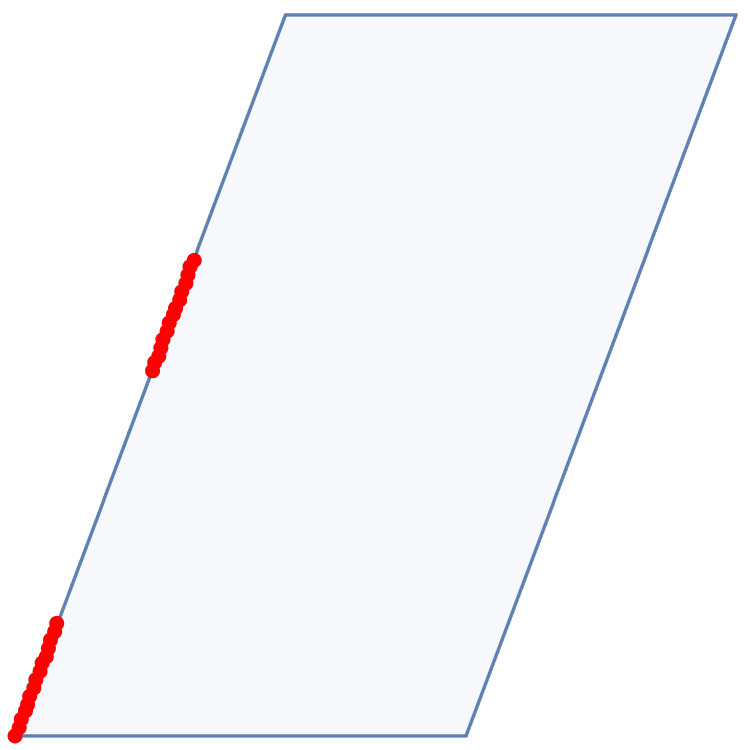}
\caption{$M = 3$}
\end{subfigure}
\caption{The eigenvalue distribution for different choices of $M$ (which are divisors of $ab$) for the basic solution with $ab = 6$, $N = 30$, drawn on a torus with periodicities $1$ and $ab\omega$. The $M=1$ case is the one analyzed in \cite{Benini:2020gjh}.}
\label{fig:different Ms}
\end{figure}

It is not clear how to evaluate $\log(Z)$ for generic choices of the $m_j$. In this section, we will generalize the result \eqref{Mequalsone} to the choice 
\begin{equation}
m_j = M\cdot j \mod ab \,,
\end{equation}
for any integer $M$. At large $N$ we expect the leading order to only depend on the (shifted) eigenvalue distribution, and one can show that this depends on $M$ only through $\gcd(M,ab)$. We will show this momentarily for the leading order in $N$. When $M=1$ the (shifted) eigenvalues are uniformly distributed along the $ab\omega$ cycle of the torus, and otherwise they are distributed as a chain of step functions along it, covering $\frac{1}{M}$ of the cycle, as in Figure \ref{fig:different Ms}.

In order to calculate $\cZ$ for the basic solution
 \eqref{eq: basic solution}, it is useful to define the
building block
\begin{equation}
\begin{aligned}
\label{eq: bare BB gen}
\tilde{\Psi}_\Delta &\equiv \sum_{i\neq j}^{N}\log\left[\tilde{\Gamma}\left(\Delta+\omega\frac{j-i}{N}+\omega\left(m_{j}-m_{i}\right);a\omega,b\omega\right)\right] \\
&= \sum_{i, j=1}^{N}\log\left[\tilde{\Gamma}\left(\Delta+\omega\frac{j-i}{N}+\omega\left(m_{j}-m_{i}\right);a\omega,b\omega\right)\right] - N\log\left[\tilde{\Gamma}\left(\Delta;a\omega,b\omega\right)\right] \\
&\equiv \Psi_\Delta - N\log\left[\tilde{\Gamma}\left(\Delta;a\omega,b\omega\right)\right]\;,
\end{aligned}
\end{equation}
which allows us to write $\cZ$ in \eqref{eq:def cZ} as
\begin{equation} \label{sumincz}
\log(\cZ) = \tilde\Psi_{\Delta_1} + \tilde\Psi_{\Delta_2} - \tilde\Psi_{\Delta_1+\Delta_2} - \tilde\Psi_{0} \;.
\end{equation}
As a result, terms that are linear or constant in $\Delta$ will cancel between the different $\tilde\Psi$'s when evaluating the index. In order to simplify the form of the building block, we utilize the identity \cite{Felder:2000}
\begin{equation}
\label{eq:elliptic gamma product identity neq tau sigma}
\tilde{\Gamma}\left(u;\tau,\sigma\right)=\prod_{r=0}^{a-1}\prod_{s=0}^{b-1}\tilde{\Gamma}\left(u+\left(r\tau+s\sigma\right);a\tau,b\sigma\right) \;,
\end{equation}
to write $\Psi_\Delta$ as
\begin{equation}
\Psi_\Delta = \sum_{r=0}^{a-1}\sum_{s=0}^{b-1} \sum_{i, j=1}^{N}\log\left[\tilde{\Gamma}\left(\Delta+\omega\frac{j-i}{N} + \omega\left(m_{j}-m_{i}+as+br\right);ab\omega,ab\omega\right)\right] \;.
\end{equation}

We can now make two simplifications. The first is to assume that $ab | N$ and denote $\tilde{N}=\frac{N}{ab}$. The justification is that \cite{Benini:2020gjh} showed that the effect of ignoring the residue of $N/(ab)$ is subleading at large $N$, when $N \gg ab$.
Let's set $i=\gamma ab+c$, $j=\delta ab+d$, with $\gamma,\delta=0,\cdots,\tilde{N}-1$
and $c,d = 1,\cdots,ab$. The building block for \begin{equation}
    m_j = M\cdot j \mod ab
\end{equation} 
will then take the form
\begin{equation}
\label{eq: BB before dropping terms}
\begin{aligned}
\Psi_\Delta &= \sum_{r=0}^{a-1}\sum_{s=0}^{b-1}\sum_{\gamma,\delta=0}^{\tilde{N}-1}\sum_{c,d=1}^{ab} \log\left[\tilde{\Gamma}\left(\Delta+\omega\frac{\delta-\gamma}{\tilde{N}} + \omega\frac{d-c}{N} + \omega\left(\eta_{dc}+as+br\right);ab\omega,ab\omega\right)\right]  \;, \\
\eta_{dc} &= \left(M d\mod ab\right) - \left(M c\mod ab\right) \;.
\end{aligned}
\end{equation}
A second simplification can be made by dropping the term $\omega\frac{d-c}{N}$, which also does not affect the result at leading order in $N$\footnote{In appendix A of \cite{Benini:2020gjh} this is shown for the case $M=1$, and with
minor modifications the proof can be generalized to any $M\in\mathbb{N}$.}.

We see that indeed at this point the $M$ dependence enters only through $\eta_{dc}$. Because\footnote{Write $M = \mu x$ where $\gcd(x,ab) = 1$ and therefore $x$ has a modular inverse modulo $ab$. Then $\sum_{c=1}^{ab} f(M c \mod ab) = \sum_{c=1}^{ab} f(\mu x c \mod ab) = \sum_{c=1}^{ab} f(\mu c \mod ab)$ and so the sum only depends on $\mu = \gcd(M,ab)$
.} of the summation over $c$, $d$, this depends only on $\gcd(M,ab)$, and so without loss of generality from now on we will assume $M=\gcd(M,ab)$.
Using $\sum_{c=1}^{ab}f\left(Mc\mod ab\right)=M\sum_{c=1}^{\frac{ab}{M}}f\left(Mc\right)$ we have
\begin{equation}
\begin{aligned}
\label{eq:before resummation reformulation}
\Psi_\Delta &= M^{2}\sum_{r=0}^{a-1}\sum_{s=0}^{b-1}\sum_{\gamma,\delta=0}^{\tilde{N}-1}\sum_{c,d=1}^{ab/M} \log\left[\tilde{\Gamma}\left(\Delta+\omega\frac{\delta-\gamma}{\tilde{N}}+\omega\left(Md-Mc+as+br\right);ab\omega,ab\omega\right)\right] \\
&= M^{2}\sum_{r=0}^{a-1}\sum_{s=0}^{b-1}\sum_{\gamma,\delta=0}^{\tilde{N}-1}\sum_{c,d=0}^{ab/M-1} \log\left[\tilde{\Gamma}\left(\Delta + \omega\frac{\delta-\gamma}{\tilde{N}} + M\omega(c+d) + \omega\left(M-ab+as+br\right);ab\omega,ab\omega\right)\right] \;,
\end{aligned}
\end{equation}
and using \eqref{eq:elliptic gamma product identity neq tau sigma} we have
\begin{equation}
\Psi_\Delta = M^{2}\sum_{r=0}^{a-1}\sum_{s=0}^{b-1}\sum_{\gamma,\delta=0}^{\tilde{N}-1} \log\left[\tilde{\Gamma}\left(\Delta + \omega\frac{\delta-\gamma}{\tilde{N}} + \omega\left(M-ab+as+br\right);M\omega,M\omega\right)\right] \;.
\end{equation}
Now we use the modular formula \eqref{eq: elliptic gamma function decomposition} to rewrite this as
\begin{equation}
\label{eq:Psi before massage}
\Psi_{\Delta} = M^2 \sum_{r=0}^{a-1}\sum_{s=0}^{b-1} \sum_{\gamma,\delta=1}^{\tilde{N}} \left[-\pi i\cQ\left(u;M\omega,M\omega\right) -\log\left[\theta_{0}\left(\frac{u}{M\omega},-\frac{1}{M\omega}\right)\right] + \sum_{k=0}^{\infty}\log\left[\frac{\psi\left(\frac{k+1+u}{M\omega}\right)}{\psi\left(\frac{k-u}{M\omega}\right)}\right]\right] \;,
\end{equation}
where 
\begin{equation}
u=\left[\Delta\right]_{\omega}+\omega\frac{\delta-\gamma}{\tilde{N}}+\omega\left(M-ab+as+br\right) \;,
\end{equation}
and the functions $\cQ$, $\theta_0$, and $\psi$ are defined in Appendix \ref{app:Special functions}. As $\Psi_\Delta$ was originally invariant under shifting $\Delta\to \Delta+1$, we shift $\Delta \to \left[\Delta\right]_{\omega}$,
which is exactly the domain where the plethystic expansion for $\theta_{0}\left(\frac{u}{\omega},-\frac{1}{\omega}\right)$ is valid, and where the sum over the $\psi$ functions converges. Note that $\left[\Delta\right]_{\omega}=\left[\Delta\right]_{M\omega}$.

\paragraph{Evaluating $\cQ$:}

After plugging in the expression for the polynomial $\cQ$ from \eqref{eq:def Q modular} and summing over $\gamma, \delta, r ,s$, the first term in \eqref{eq:Psi before massage} becomes 
\begin{multline}
    -\pi i\frac{N^{2}}{3ab\omega^{2}}B_{3}\left(\left[\Delta\right]_{\omega}+1-\frac{a+b}{2}\omega\right) \\
    - \pi i\frac{N^{2}\left(1-\frac{a^{2}}{2}-\frac{b^{2}}{2}+a^{2}b^{2}-M^{2}\right)-a^{2}b^{2}}{6ab}B_{1}\left(\left[\Delta\right]_{\omega}+1-\frac{a+b}{2}\omega\right) \;,
\end{multline}
where $B_1(x) = x - \frac{1}{2}$ and $B_3(x) = x^3 - \frac{3}{2} x^2 +\frac{1}{2}x$ are Bernoulli polynomials. The second term cancels in \eqref{sumincz}, while the first term is the same as in the $M=1$ case analyzed in \cite{Benini:2020gjh}.

\paragraph{Evaluating $\theta_0$:}
Remembering that $M|ab$, the second term in \eqref{eq:Psi before massage} is
\begin{equation}
    -M^{2}\sum_{r=0}^{a-1}\sum_{s=0}^{b-1}\sum_{\gamma,\delta=1}^{\tilde{N}}\log\left[\theta_{0}\left(\frac{\left[\Delta\right]_{\omega}}{M\omega}+\frac{\delta-\gamma}{M\tilde{N}}+\frac{as+br}{M},-\frac{1}{M\omega}\right)\right]\;.
\end{equation}
After using the Plethystic expansion \eqref{eq: plethystic definition theta} for $\theta_{0}$ it takes the form
\begin{equation}
    M^{2}\sum_{r=0}^{a-1}\sum_{s=0}^{b-1}\sum_{\gamma,\delta=1}^{\tilde{N}}\sum_{\ell=1}^{\infty}\frac{1}{\ell}\frac{\tilde{y}^{\ell}\zeta_{M\tilde{N}}^{\ell\left(\delta-\gamma\right)}\zeta_{M}^{\ell\left(as+br\right)}+\tilde{h}^{\ell}\tilde{y}^{-\ell}\zeta_{M\tilde{N}}^{\ell\left(\gamma-\delta\right)}\zeta_{M}^{-\ell\left(as+br\right)}}{1-\tilde{h}^{\ell}} \;,
\end{equation}
where
\begin{equation}
    \zeta_{v}\equiv e^{2\pi i/v}\,,\qquad\tilde{y}\equiv e^{\frac{2\pi i\left[\Delta\right]_{\omega}}{M\omega}}\,,\qquad \tilde{h}\equiv e^{-\frac{2\pi i}{M\omega}}\;,
\end{equation}
and so the second term of \eqref{eq:Psi before massage} is
\begin{multline}
        abM\sum_{\gamma,\delta=1}^{\tilde{N}}\sum_{\ell=1}^{\infty}\frac{1}{\ell}\frac{\tilde{y}^{M\ell}\zeta_{\tilde{N}}^{\ell\left(\delta-\gamma\right)}+\tilde{h}^{M\ell}\tilde{y}^{-M\ell}\zeta_{\tilde{N}}^{\ell\left(\gamma-\delta\right)}}{1-\tilde{h}^{M\ell}} = abM\tilde{N}\sum_{\ell=1}^{\infty}\frac{1}{\ell}\frac{\tilde{y}^{M\tilde{N}\ell}+\tilde{h}^{M\tilde{N}\ell}\tilde{y}^{-M\tilde{N}\ell}}{1-\tilde{h}^{M\tilde{N}\ell}} \\
        = -abM\tilde{N}\log\left[\theta_{0}\left(\frac{\tilde{N}\left[\Delta\right]_{\omega}}{\omega};-\frac{\tilde{N}}{\omega}\right)\right] \;,
\end{multline}
which is exponentially suppressed in $N$.

\paragraph{Evaluating $\psi$:}
Last but not least, we have the third term in \eqref{eq:Psi before massage}. We'll expand it using the plethystic expansion for the $\psi$ function, $\log\left[\psi\left(t\right)\right] = -\sum_{\ell=1}^{\infty}\left(\frac{t}{\ell}+\frac{1}{2\pi i\ell^{2}}\right)e^{-2\pi i\ell t}$, and then sum over $\gamma,\delta,r,s$. When $M\mid\ell$ none of the terms in the plethystic expansion vanishes. Otherwise, the second term always vanishes in the sum over $r,s$, while the first term vanishes unless $M\mid a\ell$ and we are looking at the term with $br$ multiplying the exponent, or $M\mid b\ell$ and we are looking at $as$ multiplying it. The $\ell$'s that satisfy $M\nmid\ell$ but $M\mid b\ell$ are of the form $\ell\to M\ell+\frac{M}{M_{b}}t$, with $t=1,\cdots,M_{b}-1$ and $M_{b}=\gcd\left(M,b\right)$. Similarly for $M\nmid\ell$ but $M\mid a\ell$ we have $\ell\to M\ell+\frac{M}{M_{a}}t$ with $t=1,\cdots,M_{a}-1$ and $M_{a}=\gcd\left(M,a\right)$. Thus, we can rewrite the numerator of the third term in \eqref{eq:Psi before massage} as 
\begin{multline}
\label{eq:Expanding sum of psi}
-M^{2}\sum_{r=0}^{a-1}\sum_{s=0}^{b-1}\sum_{\gamma,\delta=1}^{\tilde{N}}\sum_{k=0}^{\infty}\left[\sum_{\ell=1}^{\infty}\left(\frac{\frac{k+1}{M\omega}+\frac{\left[\Delta\right]_{\omega}}{M\omega}+\frac{\delta-\gamma}{M\tilde{N}}+1-\frac{ab}{M}+\frac{as+br}{M}}{M\ell}+\frac{1}{2\pi iM^{2}\ell^{2}}\right)e^{-2\pi iM\ell v}\right] \\
- \sum_{r=0}^{a-1}\sum_{s=0}^{b-1}\sum_{\gamma,\delta=1}^{\tilde{N}}\sum_{k=0}^{\infty}\sum_{\ell=0}^{\infty}\left[\sum_{t=1}^{M_{b}-1}\frac{a s M_b}{M_{b}\ell+t}e^{-2\pi i\left(M\ell+\frac{M}{M_{b}}t\right)v}+\sum_{t=1}^{M_{a}-1}\frac{b r M_a}{M_{a}\ell+t}e^{-2\pi i\left(M\ell+\frac{M}{M_{a}}t\right)v}\right] \;,
\end{multline}
where the first line comes from those $\ell$'s that satisfy $M\mid\ell$, while the first term in the second line comes from the $\ell$'s that satisfy $M\mid b\ell$ and $M\nmid\ell$, and the second from those that have $M\nmid\ell$ but $M\mid a\ell$, and $v$ is the argument of the $\psi$ function,
\begin{equation}
    v=\frac{k+1}{M\omega}+\frac{\left[\Delta\right]_{\omega}}{M\omega}+\frac{\delta-\gamma}{M\tilde{N}}+1-\frac{ab}{M}+\frac{as+br}{M}\ .
\end{equation}

For the first line of \eqref{eq:Expanding sum of psi} we can sum over $\delta,\gamma,r,s$ to find
\begin{multline}
-ab\sum_{k=0}^{\infty}\sum_{\ell=1}^{\infty}\left[\left(\frac{\frac{k+1}{\omega}+\frac{\left[\Delta\right]_{\omega}}{\omega}}{\ell}\tilde{N}+\frac{1}{2\pi i\ell^{2}}\right)e^{-2\pi i\tilde{N}\ell\left(\frac{k+1}{\omega}+\frac{\left[\Delta\right]_{\omega}}{\omega}\right)}\right]\\-N\left(M-\frac{a}{2}-\frac{b}{2}\right)\sum_{k=0}^{\infty}\sum_{\ell=1}^{\infty}\left[\frac{1}{\ell}e^{-2\pi i\tilde{N}\ell\left(\frac{k+1}{\omega}+\frac{\left[\Delta\right]_{\omega}}{\omega}\right)}\right] \;,
\end{multline}
and similarly, from the denominator of the third term in \eqref{eq:Psi before massage} we have 
\begin{multline}
ab\sum_{k=0}^{\infty}\sum_{\ell=1}^{\infty}\left[\left(\frac{\frac{k}{\omega}-\frac{\left[\Delta\right]_{\omega}}{\omega}}{\ell}\tilde{N}+\frac{1}{2\pi i\ell^{2}}\right)e^{-2\pi i\tilde{N}\ell\left(\frac{k}{\omega}-\frac{\left[\Delta\right]_{\omega}}{\omega}\right)}\right]\\-N\left(M-\frac{a}{2}-\frac{b}{2}\right)\sum_{k=0}^{\infty}\sum_{\ell=1}^{\infty}\left[\frac{1}{\ell}e^{-2\pi i\tilde{N}\ell\left(\frac{k}{\omega}-\frac{\left[\Delta\right]_{\omega}}{\omega}\right)}\right] \;,
\end{multline}
which are all exponentially suppressed\footnote{We note that these two infinite sums can be resummed into
\begin{equation}
ab\sum_{k=0}^{\infty}\log\left[\frac{\psi\left(\frac{\tilde{N}\left(k+1+\left[\Delta\right]_{\omega}\right)}{\omega}\right)}{\psi\left(\frac{\tilde{N}\left(k-\left[\Delta\right]_{\omega}\right)}{\omega}\right)}\right]-N\left(M-\frac{a}{2}-\frac{b}{2}\right)\log\left[\theta_{0}\left(\frac{\tilde{N}\left[\Delta\right]_{\omega}}{\omega};-\frac{\tilde{N}}{\omega}\right)\right] \;.
\end{equation}} in $N$.

We now move to the second line of \eqref{eq:Expanding sum of psi}. One uses the identities 
\begin{equation}
\sum_{n=0}^{N-1} e^{\lambda n}= \frac{1 - e^{\lambda N}}{1 - e^{\lambda}}\;,\qquad 
\sum_{n=0}^{N-1}ne^{\lambda n}=\frac{d}{d\lambda}\sum_{n=0}^{N-1}e^{\lambda n}=\frac{\left(N-1\right)e^{\lambda\left(N+1\right)}-Ne^{\lambda N}+e^{\lambda}}{\left(1-e^{\lambda}\right)^{2}}  \;,
\end{equation}
to find that 
\begin{equation}
\sum_{\gamma,\delta=1}^{\tilde{N}}e^{2\pi ix\frac{\left(\gamma-\delta\right)}{M\tilde{N}}} = \frac{\sin^{2}\left(\frac{\pi x}{M}\right)}{\sin^{2}\left(\frac{\pi x}{M\tilde{N}}\right)}     
 \;, \qquad \sum_{s=0}^{b-1}as\exp\left[\frac{2\pi i\left(M\ell+\frac{M}{M_{b}}t\right)}{M}as\right] = -\frac{ab}{1-e^{\frac{2\pi ita}{M_{b}}}} \;,
\end{equation}
which allows to sum over $r,s,\gamma,\delta$, resulting in
\begin{equation}
ab\sum_{k=0}^{\infty}\sum_{\ell=0}^{\infty}\sum_{t=1}^{M_{b}-1}\frac{a M_b}{1-e^{\frac{2\pi ita}{M_{b}}}}\frac{1}{M_{b}\ell+t}\frac{\sin^{2}\left(\frac{\pi t}{M_{b}}\right)}{\sin^{2}\left(\frac{\pi\ell}{\tilde{N}}+\frac{\pi t}{M_{b}\tilde{N}}\right)}e^{-2\pi i\left(\ell+\frac{t}{M_{b}}\right)\left(\frac{k+1}{\omega}+\frac{\left[\Delta\right]_{\omega}}{\omega}\right)}+\left(a\leftrightarrow b\right) \;.
\end{equation}
Similarly, the analogous term coming from the denominator of the third term in \eqref{eq:Psi before massage} takes the form 
\begin{equation}
ab\sum_{k=0}^{\infty}\sum_{\ell=0}^{\infty}\sum_{t=1}^{M_{b}-1}\frac{a M_b}{1-e^{-\frac{2\pi iat}{M_{b}}}}\frac{1}{M_{b}\ell+t}\frac{\sin^{2}\left(\frac{\pi t}{M_{b}}\right)}{\sin^{2}\left(\frac{\pi\ell}{\tilde{N}}+\frac{\pi t}{M_{b}\tilde{N}}\right)}e^{-2\pi i\left(\ell+\frac{t}{M_{b}}\right)\left(\frac{k}{\omega}-\frac{\left[\Delta\right]_{\omega}}{\omega}\right)}+\left(a\leftrightarrow b\right) \;.
\end{equation}
At large $N$ the sine in the denominator of these two expressions gives an $O(N^2)$ dependence. Moreover, the sum over $k$ can be easily done, leaving us with an overall contribution of the second line of \eqref{eq:Expanding sum of psi}
\begin{multline}
    ab\tilde{N}^{2}\Bigg[\sum_{\ell=0}^{\infty}\sum_{t=1}^{M_{b}-1}\frac{a}{1-e^{\frac{2\pi iat}{M_{b}}}}\frac{M_{b}^{3}\sin^{2}\left(\frac{\pi t}{M_{b}}\right)}{\pi^2\left(M_{b}\ell+t\right)^{3}} \frac{e^{2\pi i\left(\ell+\frac{t}{M_{b}}\right)\left(-\frac{1}{\omega}-\frac{\left[\Delta\right]_{\omega}}{\omega}\right)}}{1-e^{-\frac{2\pi i}{\omega}(\ell+\frac{t}{M_b})}} +\left(a\leftrightarrow b\right) \\
    + \sum_{\ell=0}^{\infty}\sum_{t=1}^{M_{b}-1}\frac{a}{1-e^{-\frac{2\pi iat}{M_{b}}}}\frac{M_{b}^{3}\sin^{2}\left(\frac{\pi t}{M_{b}}\right)}{\pi^2\left(M_{b}\ell+t\right)^{3}} \frac{e^{2\pi i\left(\ell+\frac{t}{M_b}\right)\frac{[\Delta]_\omega}{\omega}}}{1-e^{-\frac{2\pi i}{\omega}\left(\ell+\frac{t}{M_b}\right)}} + \left(a\leftrightarrow b\right)\Bigg] + O(N) \;.
\end{multline}
When $M=1$ the sum over $t$ contains no terms, so these contributions are irrelevant. We also note that in the Cardy limit, $\omega \to 0$ with fixed $a,b$, these terms are exponentially suppressed.

\paragraph{Overall contribution:}
Let's now compute $\cZ$, which to leading order in $N$ is\footnote{When computing $\lim_{\Delta\to0}\Psi_\Delta$ some of the terms in $\Psi_\Delta$ diverge. When $M=1$, these divergences directly cancel with the second term in \eqref{eq: bare BB gen}. In any case, the $O(N^2)$ terms we focus on here do not diverge, while presumably divergences in subleading orders cancel with the subleading terms we neglected.}
\begin{equation}
    \cZ = \Psi_{\Delta_1} + \Psi_{\Delta_2} - \Psi_{\Delta_1 + \Delta_2} - \Psi_0 + O(N) \;.
\end{equation}
The chemical potentials satisfy either $[\Delta_1+\Delta_2]_\omega = [\Delta_1]_\omega + [\Delta_2]_\omega$, or $[\Delta_1+\Delta_2]_\omega = [\Delta_1]_\omega + [\Delta_2]_\omega + 1$. For the first case,
\begin{multline}
\label{eq:large N contrib 1st case}
    \cZ = -\pi i N^2 \frac{[\Delta_1]_\omega [\Delta_2]_\omega [\Delta_3]_\omega}{\tau \sigma} \\
    + \frac{N^{2}}{ab}\sum_{\Delta} \eta_\Delta \Bigg[\sum_{\ell=0}^{\infty}\sum_{t=1}^{M_{b}-1}\frac{a}{1-e^{\frac{2\pi iat}{M_{b}}}}\frac{M_{b}^{3}\sin^{2}\left(\frac{\pi t}{M_{b}}\right)}{\pi^2\left(M_{b}\ell+t\right)^{3}} \frac{e^{2\pi i\left(\ell+\frac{t}{M_{b}}\right)\left(-\frac{1}{\omega}-\frac{\Delta}{\omega}\right)}}{1-e^{-\frac{2\pi i}{\omega}(\ell+\frac{t}{M_b})}} +\left(a\leftrightarrow b\right) \\
    + \sum_{\ell=0}^{\infty}\sum_{t=1}^{M_{b}-1}\frac{a}{1-e^{-\frac{2\pi iat}{M_{b}}}}\frac{M_{b}^{3}\sin^{2}\left(\frac{\pi t}{M_{b}}\right)}{\pi^2\left(M_{b}\ell+t\right)^{3}} \frac{e^{2\pi i\left(\ell+\frac{t}{M_b}\right)\frac{\Delta}{\omega}}}{1-e^{-\frac{2\pi i}{\omega}\left(\ell+\frac{t}{M_b}\right)}} + \left(a\leftrightarrow b\right)\Bigg] \;,
\end{multline}
where $[\Delta_3]_\omega \equiv \tau + \sigma - [\Delta_1]_\omega - [\Delta_2]_\omega - 1$, the sum $\sum_\Delta$ is over ${\Delta \in \{[\Delta_1]_\omega, [\Delta_2]_\omega, [\Delta_1 + \Delta_2]_\omega, 0\}}$, and we define $\eta_\Delta = \{1,1,-1,-1\}$ respectively. For the second case a similar formula applies\footnote{We neglected an imaginary part that determines the phase of the contribution to the index, which in any case is sensitive to $O(1)$ terms.} using the function $[\Delta]^\prime_\omega = [\Delta]_\omega + 1$,
\begin{multline}
\label{eq:large N contrib 2nd case}
    \cZ = -\pi i N^2 \frac{[\Delta_1]_\omega^\prime [\Delta_2]_\omega^\prime [\Delta_3]_\omega^\prime}{\tau \sigma} \\
    + \frac{N^{2}}{ab}\sum_{\Delta} \eta_\Delta \Bigg[\sum_{\ell=0}^{\infty}\sum_{t=1}^{M_{b}-1}\frac{a}{1-e^{\frac{2\pi iat}{M_{b}}}}\frac{M_{b}^{3}\sin^{2}\left(\frac{\pi t}{M_{b}}\right)}{\pi^2\left(M_{b}\ell+t\right)^{3}}  \frac{e^{2\pi i\left(\ell+\frac{t}{M_b}\right)\frac{\Delta}{\omega}}}{1-e^{-\frac{2\pi i}{\omega}\left(\ell+\frac{t}{M_b}\right)}}+\left(a\leftrightarrow b\right) \\
    + \sum_{\ell=0}^{\infty}\sum_{t=1}^{M_{b}-1}\frac{a}{1-e^{-\frac{2\pi iat}{M_{b}}}}\frac{M_{b}^{3}\sin^{2}\left(\frac{\pi t}{M_{b}}\right)}{\pi^2\left(M_{b}\ell+t\right)^{3}} \frac{e^{2\pi i\left(\ell+\frac{t}{M_{b}}\right)\left(-\frac{1}{\omega}-\frac{\Delta}{\omega}\right)}}{1-e^{-\frac{2\pi i}{\omega}(\ell+\frac{t}{M_b})}} + \left(a\leftrightarrow b\right)\Bigg] \;,
\end{multline}
where now $[\Delta_3]_\omega^\prime \equiv \tau + \sigma - [\Delta_1]_\omega^\prime - [\Delta_2]_\omega^\prime + 1$, the sum $\sum_\Delta$ is over ${\Delta \in \{[\Delta_1]_\omega^\prime, [\Delta_2]_\omega^\prime, [\Delta_1 + \Delta_2]_\omega^\prime, 0\}}$, and we define $\eta_\Delta = \{1,1,-1,-1\}$ as before.

\subsection{Dominance of new contributions}
Are these new contributions to the index always subleading with respect to the $M = 1$ contribution analyzed in \cite{Benini:2020gjh} (and the other contribution analyzed in \cite{Colombo:2021kbb} which is equal to it in the large $N$ limit)? The answer turns out to be negative. We can verify numerically that there are choices of parameters, say $\omega = 0.45 + 0.84 i$, $\Delta_1 = 0.07 + 0.13 i$, $\Delta_2 = 0.02 + 0.11 i$, $a = 2$, $b = 3$, $N = 300$, where the $M = 1$ contribution is smaller then, say, the $M = 6$ contribution. This is true both for our approximate large $N$ value in \eqref{eq:large N contrib 1st case}-\eqref{eq:large N contrib 2nd case} and for the exact evaluation \eqref{sumincz}.

Thus, in some cases this term may be more dominant. Note however,
that there may exist \emph{even more dominant }terms for other types
of $\left\{ m_{j}\right\} $ shifts, that we have not yet computed. Moreover, this $\omega$ is not necessarily in the regime where this Bethe Ansatz solution is the most dominant one, and so these contributions might be cancelled by contributions from other solutions to the Bethe Ansatz equations. Analyzing these cancellations requires knowledge both about the phase of $\cI_u$, which is sensitive to $O(1)$ terms in the expansion, and about other $\left\{ m_{j}\right\} $ shifts, whose analysis is beyond the scope of this work.

One might wonder what happens in the Cardy limit, $\omega \to 0$, where the contribution from this solution is usually the dominant one to the index. In that limit one can show that the new, $M$-dependent, terms at large $N$ in \eqref{eq:large N contrib 1st case} become exponentially suppressed in $1/\omega$, and so the new contributions are similar to the $M=1$ case.

\section{The Bethe Ansatz for $SU(2)$}
\label{sec:su2}


In this section we discuss the special case of $SU(2)$ gauge group, for which we analyzed in more detail the behavior of the index for large values of $a$ and $b$, and in particular the  
case of $b=a+1$ (which in the limit of large $a,b$ should converge to the results for equal chemical potentials). In this case the solutions to the Bethe ansatz equation are fully classified ($u_{12}=\frac{1}{2}, \frac{\omega}{2}, \frac{1}{2}+\frac{\omega}{2}$), and there is a single variable $u=u_{12}$ (with $u_1=-u_2$), and a single integer $m$ labelling the different contributions (for each solution to the Bethe ansatz). One thing we will show is that the contribution to the index from some values of $m$
is exponentially large in $a$ for large $a$, such that large cancellations between different contributions must occur in the large $a$ limit.


\subsection{Moving $Z_{m}^{u}$ by $n\tau$}

We assume without loss of generality that $b>a$. 
For $SU(2)$ we have \eqref{eq:def cZ}
\begin{equation}
\mathcal{Z}\left(u ; \Delta,\sigma,\tau\right) = 
\frac{\tilde{\Gamma}\left(u+\Delta_{1};\tau,\sigma\right)\tilde{\Gamma}\left(u+\Delta_{2};\tau,\sigma\right)}{\tilde{\Gamma}\left(u;\tau,\sigma\right)\tilde{\Gamma}\left(u+\Delta_{1}+\Delta_{2};\tau,\sigma\right)}
\frac{\tilde{\Gamma}\left(-u+\Delta_{1};\tau,\sigma\right)\tilde{\Gamma}\left(-u+\Delta_{2};\tau,\sigma\right)}{\tilde{\Gamma}\left(-u;\tau,\sigma\right)\tilde{\Gamma}\left(-u+\Delta_{1}+\Delta_{2};\tau,\sigma\right)}.
\end{equation}
We will denote $Z_m^u = \mathcal{Z}\left(u-m\omega;\Delta,\sigma,\tau\right)$, such that the contribution of each solution $u$ of the BA equation to the index is proportional to $\sum_{m=1}^{ab} Z_m^u$. This includes both the $m$-shifts and the $\beta$-shifts discussed above, since we showed in section \ref{sec:Reduced BA} that together they give $\sum_{m=1}^{2ab} Z_m^u = 2 \sum_{m=1}^{ab} Z_m^u$.

Before beginning the analysis we derive a formula for how the contribution $Z_m^u$ changes under shifts of $m$.
Using \eqref{eq:elliptic gamma quasi periodicity} we have
\begin{equation}
    \begin{aligned}
        Z_{m+na}^u&=\mathcal{Z}\left(u-m\omega-na\omega;\Delta,\sigma,\tau\right)=\mathcal{Z}\left(u-m\omega-n\tau;\Delta,\sigma,\tau\right) \\
        &= \mathcal{Z}\left(u-m\omega;\Delta,\sigma,\tau\right) \\
        &\quad\times \prod_{l=1}^{n}\frac{\theta_{0}\left(u-m\omega-\left(al\mod b\right)\omega,\sigma\right)\theta_{0}\left(\Delta_{1}+\Delta_{2}+u-m\omega-\left(al\mod b\right)\omega,\sigma\right)}{\theta_{0}\left(\Delta_{1}+u-m\omega-\left(al\mod b\right)\omega,\sigma\right)\theta_{0}\left(\Delta_{2}+u-m\omega-\left(al\mod b\right)\omega,\sigma\right)} \\
        &\quad \times \prod_{l=0}^{n-1}\frac{\theta_{0}\left(\Delta_{1}-u+m\omega+\left(al\mod b\right)\omega,\sigma\right)\theta_{0}\left(\Delta_{2}-u+m\omega+\left(al\mod b\right)\omega,\sigma\right)}{\theta_{0}\left(-u+m\omega+\left(al\mod b\right)\omega,\sigma\right)\theta_{0}\left(\Delta_{1}+\Delta_{2}-u+m\omega+\left(al\mod b\right)\omega,\sigma\right)}.\label{eq:moving z by ntau}
    \end{aligned}
\end{equation}
Since $a$ and $b$ are mutually prime, for $n\leq b$ we will
get $n$ distinct values of $(al\mod b)$ in each product.

Let us look at the following $a$ evaluations of $\mathcal{Z}$:
\begin{equation}
Z_{m-a}^{u},\ \qquad \ m=1,\cdots,a,\label{eq:basic window}
\end{equation}
and call these points our ``basic window''. Even though these values
do not enter the sum in the index, we can construct from them all the terms that
do enter the sum. To get $Z_{m}^{u}$ using (\ref{eq:moving z by ntau}),
we need to take
\begin{multline}
Z_{m}^{u} = Z_{\left(m-a\right)+a}^{u}=\mathcal{Z}\left(u-\left(m-a\right)\omega-a\omega;\Delta,\sigma,\tau\right) \\
 \qquad = Z_{m-a}^{u}\frac{\theta_{0}\left(u-m\omega,\sigma\right)\theta_{0}\left(\Delta_{1}+\Delta_{2}+u-m\omega,\sigma\right)}{\theta_{0}\left(\Delta_{1}+u-m\omega,\sigma\right)\theta_{0}\left(\Delta_{2}+u-m\omega,\sigma\right)} \\ 
\qquad \qquad \times \frac{\theta_{0}\left(\Delta_{1}-u+\left(m-a\right)\omega,\sigma\right)\theta_{0}\left(\Delta_{2}-u+\left(m-a\right)\omega,\sigma\right)}{\theta_{0}\left(-u+\left(m-a\right)\omega,\sigma\right)\theta_{0}\left(\Delta_{1}+\Delta_{2}-u+\left(m-a\right)\omega,\sigma\right)}.
\end{multline}
In general, following (\ref{eq:moving z by ntau}), to construct $Z_{(m-a)+na}^{u}$
we take
\begin{equation}
\begin{aligned}
Z_{\left(m-a\right)+\left(n-1\right)a}^{u}& \frac{\theta_{0}\left(u-\left(m-a\right)\omega-\left(an\mod b\right)\omega,\sigma\right)}{\theta_{0}\left(\Delta_{1}+u-\left(m-a\right)\omega-\left(an\mod b\right)\omega,\sigma\right)} \\
&\times \frac{\theta_{0}\left(\Delta_{1}+\Delta_{2}+u-\left(m-a\right)\omega-\left(an\mod b\right)\omega,\sigma\right)}{\theta_{0}\left(\Delta_{2}+u-\left(m-a\right)\omega-\left(an\mod b\right)\omega,\sigma\right)} \\
&\times \frac{\theta_{0}\left(\Delta_{1}-u+\left(m-a\right)\omega+\left(a\left(n-1\right)\mod b\right)\omega,\sigma\right)}{\theta_{0}\left(-u+\left(m-a\right)\omega+\left(a\left(n-1\right)\mod b\right)\omega,\sigma\right)} \\ 
&\times \frac{\theta_{0}\left(\Delta_{2}-u+\left(m-a\right)\omega+\left(a\left(n-1\right)\mod b\right)\omega,\sigma\right)}{\theta_{0}\left(\Delta_{1}+\Delta_{2}-u+\left(m-a\right)\omega+\left(a\left(n-1\right)\mod b\right)\omega,\sigma\right)}.
\end{aligned}
\end{equation}
We get a recursive expression for $Z_{\left(m-a\right)+na}^{u}$ in
terms of $Z_{\left(m-a\right)+\left(n-1\right)a}^{u}$ and one of
$b$ possible factors, labeled by the distinct label $(an\mod b)$. Using it
multiple times will give us an expression for $Z_{\left(m-a\right)+na}^{u}$
in terms of $Z_{m-a}^{u}$ from the basic window, multiplied by $n$ factors
all labeled with distinct labels $(al\mod b)$, $l=1,\cdots,n$ (if $n\leq b$).
This is enough motivation to define the factors
\begin{equation}
    \begin{aligned}
    \label{eq:theta factor}
        \Theta_{r}^{u} &\equiv \frac{\theta_{0}\left(u-\left(1-a\right)\omega-r\omega,\sigma\right)\theta_{0}\left(\Delta_{1}+\Delta_{2}+u-\left(1-a\right)\omega-r\omega,\sigma\right)}{\theta_{0}\left(\Delta_{1}+u-\left(1-a\right)\omega-r\omega,\sigma\right)\theta_{0}\left(\Delta_{2}+u-\left(1-a\right)\omega-r\omega,\sigma\right)} \\
        &\qquad \times \frac{\theta_{0}\left(\Delta_{1}-u+\left(1-a\right)\omega+\left(\left(r-a\right)\mod b\right)\omega,\sigma\right)}{\theta_{0}\left(-u+\left(1-a\right)\omega+\left(\left(r-a\right)\mod b\right)\omega,\sigma\right)} \\
        &\qquad \times \frac{\theta_{0}\left(\Delta_{2}-u+\left(1-a\right)\omega+\left(\left(r-a\right)\mod b\right)\omega,\sigma\right)}{\theta_{0}\left(\Delta_{1}+\Delta_{2}-u+\left(1-a\right)\omega+\left(\left(r-a\right)\mod b\right)\omega,\sigma\right)} \\
        &= \frac{\theta_{0}\left(u-\left(1+r-a\right)\omega,\sigma\right)\theta_{0}\left(\Delta_{1}+\Delta_{2}+u-\left(1+r-a\right)\omega,\sigma\right)}{\theta_{0}\left(\Delta_{1}+u-\left(1+r-a\right)\omega,\sigma\right)\theta_{0}\left(\Delta_{2}+u-\left(1+r-a\right)\omega,\sigma\right)} \\
        & \qquad \times \frac{\theta_{0}\left(\Delta_{1}-u+\left(1+r-2a\right)\omega,\sigma\right)\theta_{0}\left(\Delta_{2}-u+\left(1+r-2a\right)\omega,\sigma\right)}{\theta_{0}\left(-u+\left(1+r-2a\right)\omega,\sigma\right)\theta_{0}\left(\Delta_{1}+\Delta_{2}-u+\left(1+r-2a\right)\omega,\sigma\right)}.
    \end{aligned}
\end{equation}
It is easy to check
that $\Theta_{r}^{u}$ are
invariant under $r\rightarrow r+b\Leftrightarrow u\rightarrow u+\sigma$,
such that the last equality is valid, and indeed we have $b$ different
factors and not more. Note that one can also shift $Z_{(m+1-a)}^{u}$
to $Z_{(m+1-a)+na}^{u}$ using only the same factors $\Theta_{r}^{u}$,
and so on. So if we calculate all of the $b$ factors of $\Theta_{r}^{u}$
($r=0,\cdots,b-1$), we get all the factors needed to produce all the terms
in the sum out of our basic window values (\ref{eq:basic window}).
If we arrange the $\Theta_{r}^{u}$ in the specific order
\begin{equation}
\left(\Theta_{na\mod b}^{u}\right)_{n=1}^{b},
\end{equation}
then we get the ordered factors needed to move $Z_{1-a}^{u}$ to $Z_{ab+1-a}^{u}$,
through all $Z_{na+1-a}^{u}$ values. Using 
\begin{equation}
\left(\Theta_{\left(na+1\right)\mod b}^{u}\right)_{n=1}^{b}
\end{equation}
we get the ordered factors needed to move $Z_{2-a}^{u}$ to $Z_{ab+2-a}^{u}$,
passing through all $Z_{na+2-a}^{u}$ values, and so on.

\begin{figure}[t]
\centering
\includegraphics{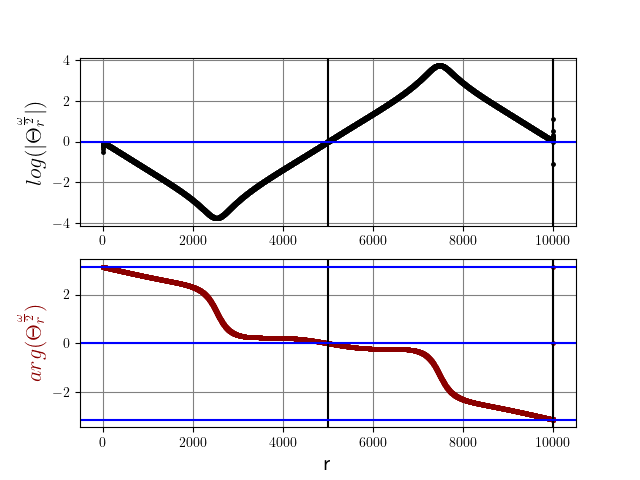}
\caption{We plot $\Theta_{r}^{u=\frac{\omega}{2}}$ for $a=9999,\ b=10000$,
$\tau=-0.67+2.3i$, $\Delta_{1}=0.3+1.1i$ and $\Delta_{2}=0.11-0.6i$.
The black vertical lines show two points in which $|\Theta_{r}^{u}|=1$,
that represent the beginning and end of a streak of $|\Theta_{r}^{u}|$
values that are larger than $1$. The blue horizontal lines mark the
special values of $0$ for $\log\left(|\Theta_{r}^{\frac{\omega}{2}}|\right)$,
and $-\pi,0,\pi$ for $\arg\left(\Theta_{r}^{\frac{\omega}{2}}\right).$}
\label{fig:trial3}
\end{figure}

Computing these factors numerically for generic parameters, their absolute values and phases
generally look something like Figure \ref{fig:trial3} (this is for the solution $u=\frac{\omega}{2}$, we will discuss the other solutions later). It should be noted that for some parameters, the $|\Theta_{r}^{u}|$
graph looks a bit different. It can cross the $|\Theta_{r}^{u}|=0$
line more often, and it can be relatively flat. With minor changes,
the arguments below will hold also in these cases.

When we take the large $a$ limit, $\Theta_r^u$ approaches a continuous function of $(r/b)$ (as in figure \ref{fig:trial3}).
Naively one may think that since when we take $a, b \to \infty$ with fixed $\tau$, $\omega$ goes to zero, one may be able to approximate the sum over $m$ as a continuous integral over $x=m/a$ (in the range $[0,b]$). However, using the fact that most $\Theta_r^u$ are very different from one (even for large $a$), one can show that even if two adjacent values in our ``basic window'' approach each other in the large $a$ limit, this is not true once they are shifted by a large amount (so that $x$ is of order $a$). So $Z_m^u$ does not really approach a continuous function of $x$ in the large $a$ limit.

\subsubsection{The $b=a+1$ Case}

Consider now the special case $b=a+1$, for which $(na\mod b)=(\left(-n\right)\mod b)$.
Then, the order of the factors that we multiply by will be just as plotted in Figure \ref{fig:trial3} but reversed,
with the specific starting point determined by $m$. This case is particularly
interesting because for fixed $\tau$, the limit $a\rightarrow\infty$
is equivalent to $\tau\rightarrow\sigma$, where we expect to converge to the results for equal chemical potentials $a=b=1$ (which are given by \eqref{eq: BA formulation sum-1} with no sum over $m$).

Looking at Figure \ref{fig:trial3}, we see that (\ref{eq:theta factor}) is roughly
divided to a continuous half of $\Theta_r^u$'s with absolute value greater
than 1, and half which are smaller than 1. It can also easily
be shown that the factors in each half are just the inverses of the values
in the other half, with a mirrored order (this is precisely true when $u=\frac{\omega}{2}$
and $\frac{1}{2}+\frac{\omega}{2}$, and true for large $a$ for $u=\frac{1}{2}$). Note that the shape
of the figure does not change when taking $a\rightarrow\infty,\ \tau=const$,
it just becomes denser (except at the edges of the graph,
in a region that becomes negligible as $a\rightarrow\infty$). The
same is true for the basic window values of $Z_{m-a}^{u}$.

\begin{figure}[t]
\includegraphics{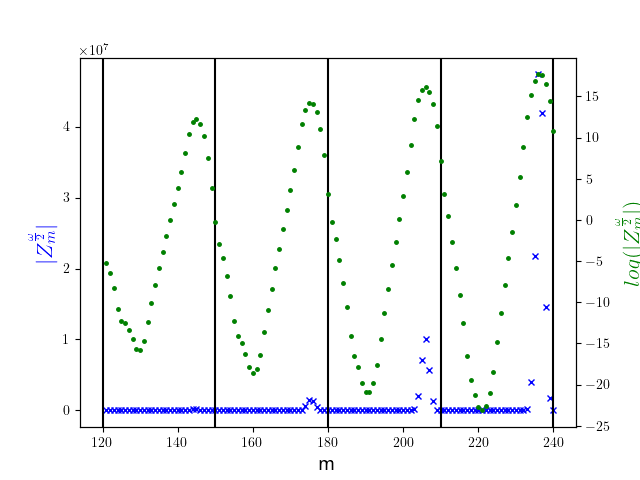}
\caption{We plot $|Z_{m}^{\frac{\omega}{2}}|$ (blue) and $\log(|Z_{m}^{\frac{\omega}{2}}|)$
(green) as a function of $m$. The horizontal lines mark the beginning
and end of $a$-point windows (here $a=30,\ b=31$, and the rest of the
parameters are as in Figure \ref{fig:trial3}). The values of each window can be calculated
from the values of the former window by multiplying its values by
$\Theta_{r}^{\frac{\omega}{2}}$ with the correct shift.}
\label{fig:z9}
\end{figure}

Moving the basic window of points by $a$ to $Z_{m}^{u},\ m=1,\cdots,a$,
each point will get multiplied by an appropriate factor from $\Theta_{b-1}^{u},\Theta_{0}^{u},\Theta_{1}^{u},\cdots,\Theta_{b-3}^{u}$, in the order they are written
($\Theta_{b-2}^{u}$ doesn't take part). Moving the new (moved) window again will result in multiplying
it by the same factors in the same order, but moved by one slot (so this
time $\Theta_{b-3}^{u}$ doesn't take part). Most of the points that
were multiplied by factors with absolute value larger than 1 before,
will get this treatment again because of the topography of $\Theta_{r}^{u}$.
Moving a few windows forward from the basic window (when $b$ is large
we have plenty), less and less points have only been multiplied by
factors with absolute value larger than 1, and more and more get multiplied
by mixed factors. So after a while, a peak in the absolute value of $Z_{m}^{u}$
arises in the region of the window that can be traced to the basic
window only through factors with absolute value greater than 1. This is demonstrated in Figure \ref{fig:z9}.
This shows us that the points in which $|\Theta_{r}^{u}|=1$ are important.
We can find these points by noting that
\begin{equation}
\Theta_{r}^{u}=\Theta_{0}^{u-r\omega}\equiv\Theta_{0}^{\tilde{u}}\underset{a\rightarrow\infty}{\longrightarrow}-Q\left(\tilde{u};\Delta,\tau\right),\label{eq:theta equals q}
\end{equation}
so finding these points means solving 
\begin{equation}
|Q\left(\tilde{u};\Delta,\tau\right)|=1.\label{eq:streaks equation}
\end{equation}
 This is similar to the BAE, except that $Q$ can have a phase different from
$0$ or $\pi$, and we only search for solutions of the form $\alpha\omega$
or $\frac{1}{2}+\alpha\omega$ for $\alpha\in\mathbb{R}$. The known BAE solutions immediately
provide us with the solutions $\tilde{u}=0,\frac{\tau}{2},\tau$
for $u=\frac{\omega}{2}$, and $\tilde{u}=\frac{1}{2},\frac{1}{2}+\frac{\tau}{2},\frac{1}{2}+\tau$
for $u=\frac{1}{2}$ and $u=\frac{1}{2}+\frac{\omega}{2}$. It is
important to remember that in most of these points $-Q(\tilde{u})=1$,
except for $\tilde{u}=0,\tau$, in which $-Q\left(\tilde{u}\right)=-1$.

We can interpret Figure \ref{fig:trial3} in light of this observation. We indeed
have $|\Theta_{r}^{u}|=1$ at $r\omega\approx0,\frac{\tau}{2},\tau$,
with the phases we expect.

\subsection{There are large cancellations between different $m$-shifted terms\label{subsec:There-are-Large large Zs}}

The fact that many $\Theta_r^u$'s are larger than $1$ implies that for large $a$ some $Z$'s, arising from multiplying many of these $\Theta$'s, will be exponentially large. One may wonder if perhaps the sum over $m$ could be dominated by some specific large values, such that we can neglect the rest of the contributions. Clearly this is impossible, given that the large $a$ limit of the sum should be a constant (namely the index for $\sigma=\tau$).
It turns out that indeed the largest contributions partially cancel each other, with many terms taking part in this process. What
is left from this partial cancellation can still be very large, but
comparable to other terms in the sum.

The picture of Figure \ref{fig:z9}, in which there is a peak in each window, with an increasing absolute value for the peak
in each window, which arises because each window is multiplied by $\Theta_{r}^{u}$
factors with absolute value larger than 1, ends at $m\approx\frac{ab}{2}$.
After that point, $\Theta_{r}^{u}$ becomes smaller than $1$.
The maximal point in the sum over shifts by $na$ (for large enough $a$) will be the one
that comes from the basic window by getting multiplied by all $\Theta_{r}^{u}$
factors in the range of values of $r$ that obeys $|\Theta_r^u| > 1$,
starting and ending with $|\Theta_{r}^{u}|$ values near $1$. 
Note that the values near the beginning and end of this range are
very close to $1$, in fact infinitesimally close for $a\rightarrow\infty$.
So around the maximum we have
many other points that are very close to it in absolute value. But if we look at
the phase of $\Theta_r^u$ near the beginning and end of the streak in
Figure \ref{fig:trial3}, it is approximately $\pi$. So we have many large terms summed
up with almost opposite phases,
that cancel each others' contributions in a noisy manner.
As we look further from the actual maximum, the change in the absolute value
between adjacent terms grows, and the relative phase also slowly changes
from $\pi$. But these terms still partially cancel each other, until
we move enough to continuously shift to a different regime. So what
is left after summing all these large terms is unclear, and can still
be very large.


The important feature for these cancellations is that either
at the beginning or at the end of the large $|\Theta_{r}^{u}|$ streak,
the phase of $\Theta_{r}^{u}$ is not $0$. Thus, the only cases that will
not have this kind of cancellations are those in which $\Theta_{r}^{u}=1$
in the beginning and end of the region, meaning when it begins and
ends with $\tilde{u}$ that solves the reduced BAE. As mentioned above, this is the case
with the other two solutions $u=\frac{1}{2},\,\frac{\omega}{2}+\frac{1}{2}$,
where the cancellations happen differently.

In these cases, it turns out that the cancellations happen between the two different solutions, rather than between different $m$-movements of the same BA solution. In the large $a$ limit, in the vicinity of the largest $|Z_{m}^{u}|$, the values of $Z_{m}^{u}$
for many different nearby $m$ terms are very similar; this is because they begin from nearby values in the ``basic window'', which are all then multiplied by almost all values of $\Theta_r^u$ that obey $|\Theta_r^u| > 1$. But if $Z_{m}^{\frac{1}{2}}$
is very close to $Z_{m+1}^{\frac{1}{2}}$, then they are both close
to $Z_{m}^{\frac{\omega}{2}+\frac{1}{2}}=Z_{m+\frac{1}{2}}^{\frac{1}{2}}$,
because $\mathcal{Z}$ is continuous. So the two solutions, $u=\frac{1}{2}$ and $u=\frac{\omega}{2}+\frac{1}{2}$, produce
very similar values of $Z_{m}^{u}$ for the largest $|Z_{m}^{u}|$
terms. Numerically we find that $H^{\frac{1}{2}}\approx-H^{\frac{\omega}{2}+\frac{1}{2}}$
at large $a$ and (at least at) small $q$, so these large contributions mostly cancel
each other. Note that if there are some extra solutions
to (\ref{eq:streaks equation}) that break the streak with non-zero phase,
it will just mean that the large contributions will cancel for the original
reason, and not that they don't cancel.

Following the last note, the $\Theta_{r}^{u}$ picture is not always
simple, depending on the precise parameters. Sometimes its absolute value crosses
$1$ multiple times as discussed above, and we get several large $|\Theta_{r}^{u}|$
streaks, and several small $|\Theta_{r}^{u}|$ streaks. This does
not change the overall picture. When taking large $a$ these regions
become denser, and so some points in the sum become exponentially
large in $a$. In this case it is not so easy to know what the global
maximum is, since it depends on the sizes of the different large $|\Theta_{r}^{u}|$
streaks and on the value of the factors in these streaks. But everything
we argued before will still be true for the global maximum in these cases,
including the partial cancellation.

The discussion of this subsection is not a feature just of $b=a+1$. It is true for any $b=a+const$
when taking large $a$, except that these cases will have more $|\Theta_{r}^{u}|=1$
crossings. 


Numerical computations of (\ref{eq: BA formulation sum-1}) 
are consistent with the above discussion. The contribution of the sum over
$m$-shifted solutions originating from $u=\frac{\omega}{2}$ partially
cancels within itself. This can be seen from the fact that the whole
sum is smaller in absolute value than the maximal term in the sum,
sometimes by orders of magnitude.

The sums originating from $u=\frac{1}{2},\frac{1}{2}+\frac{\omega}{2}$
are sometimes larger than their maximal value, but they partially
cancel each other. The rest of the cancellation comes from adding the
$\frac{\omega}{2}$ contribution, to get a value that is not exponentially
large in $a$. 

One could have thought that in the $a\rightarrow\infty$
limit, each HL solution to $Q(u;\Delta,\omega)=-1$ (including
its $m$ and $\beta$ shifts) will contribute exactly the value that
the solution to $Q\left(u;\Delta,\tau\right)=-1$ contributes at $\sigma=\tau$,
matching the solutions via $\omega\leftrightarrow\tau$. However, we see that
this is not true, as 
the contributions of some solutions to
$Q\left(u;\Delta,\omega\right)=-1$ to the index grow exponentially for large
$a$. Thus, the mapping between BA solutions and gravitational solutions described in \cite{Aharony:2021zkr} needs to be modified for this case of unequal angular momentum potentials.

\section{Allowed shifts from the gravity side}
\label{sec:shifts from gravity}

While most of this paper concentrated on some specific contributions to the index at $O(N^2)$, in this section we consider the mapping in the opposite direction, and we argue that when $\tau \neq \sigma$ there may be some additional gravitational contributions whose origin within the Bethe ansatz formalism is unknown at the moment. In order to analyze the different backgrounds, remember that the chemical potentials on the gravity side are the same as those on the CFT up to an integer shift \cite{Aharony:2021zkr}
\begin{equation}
\Delta_{g,1}=\Delta_{1}+n_{1}, \quad \Delta_{g,2}=\Delta_{2}+n_{2},\quad \tau_{g}=\tau+k_{2},\quad \sigma_{g}=\sigma+k_{1} \;.
\end{equation}
On the gravity side the on-shell action (on the first branch) takes the form 
\begin{equation}
    I = \pi iN^{2}\frac{\Delta_{g,1}\Delta_{g,2}\Delta_{g,3}}{\tau_{g}\sigma_{g}} \;,
\end{equation}
where $\Delta_{g,3}=\tau_{g}+\sigma_{g}-\Delta_{g,1}-\Delta_{g,2}-1$. The entire partition function is periodic under integer shifts of any of the four chemical potentials, while tuning $\Delta_{g,3}$ to preserve this linear relation. However, the on-shell action is not periodic. In \cite{Aharony:2021zkr} the periodicity was understood as coming from the contribution of different bulk geometries. The case where $\tau = \sigma$ was considered, and some of the different bulk geometries were associated with the shift $(\tau_g, \sigma_g) \to (\tau_g + 1, \sigma_g + 1)$. These were then argued to be matched with the contribution of the Hong-Liu solutions \eqref{eq:HL solutions} with different $r$'s.

Ostensibly, there could be also shifts in the $\Delta_g$'s, which do not seem to match to any Hong-Liu solution. However, it turned out \cite{Aharony:2021zkr} that the resulting bulks were all unstable to brane nucleation, such that the stable bulks matched in a one-to-one fashion with the contributions coming from the Bethe Ansatz solutions. The brane involved is a Euclidean D3-brane which wraps an $S^1 \subset AdS_5$ and an $S^3 \subset S^5$, and has an action which is one of
\begin{equation}
    I_{D_{3}}=\begin{cases}
2\pi N\frac{\Delta_{g,i}}{\tau_{g}}\\
2\pi N\frac{\Delta_{g,i}}{\sigma_{g}}
\end{cases}
\end{equation}
depending on the exact cycles that the brane wraps around, see details in \cite{Aharony:2021zkr}. Since the branes do not wrap the thermal cycle, their contribution to the Euclidean partition function is $e^{iI_{D_{3}}}$, and so the geometry is stable only if
\begin{equation}
\im\left(I_{D_{3}}\right) \geq 0\;.
\end{equation}

We will now repeat this analysis for the case where $\tau=a\omega$ and $\sigma = b\omega$, so that we have $\tau_{g}=a\omega+k_{1}$ and $\sigma_{g}=b\omega+k_{2}$. We will consider only shifts of $\tau_g$ and $\sigma_g$, so we consider stability bounds that are independent of the $\Delta_g$'s. The stability conditions coming from combinations of three D3-branes that wrap the same $S^{1}\subset AdS_{5}$ and three different choices of $S^{3}\subset S^{5}$ are:
\begin{equation}
    \im\left(\frac{\tau_{g}-1}{\sigma_{g}}\right) \geq 0 \,, \qquad \im\left(\frac{\sigma_{g}-1}{\tau_{g}}\right) \geq 0 \;.
\end{equation}
This implies\footnote{Note that the inequality is independent of $\re(\omega)$, as it cancels between the two terms of the second inequality.}
\begin{equation} \label{bounds}
\begin{gathered}
\im(\tau_{g}) > \im(\tau_{g})\re(\sigma_{g})-\re(\tau_{g})\im(\sigma_{g}) > -\im(\sigma_{g}) \;,\\
a > a\re(\sigma_{g})-b\re(\tau_{g}) > -b \;, \\
a > a k_2 - b k_1 > -b \;.
\end{gathered}
\end{equation}
Starting from a stable bulk solution, shifting $\tau_g\to \tau+a$ and $\sigma_g\to \sigma_g + b$ is always allowed. These shifts keep $\frac{\tau_g}{\sigma_g} = \frac{a}{b}$, and correspond to $\omega\to\omega+1$. They reproduce the Hong-Liu solutions with different $r$, \eqref{eq:HL solutions}. However, \eqref{bounds} can have other solutions. For example, when $(a,b) = (2,3)$ we can also choose $k_2 = 2$, $k_1=1$, and this shift is not reproduced by merely considering the different Hong-Liu solutions. Conceivably, these other shifts might be reproduced from the sum over the different $\{m_j\}$ shifts considered in the rest of this paper, but this is beyond the scope of this paper.

\paragraph{Acknowledgements}
We would like to thank Francesco Benini for useful discussions.
This work was supported in part by an Israel Science Foundation (ISF) center for excellence grant (grant number 2289/18), by ISF grant no. 2159/22, by Simons Foundation grant 994296 (Simons
Collaboration on Confinement and QCD Strings), by grant no. 2018068 from the United States-Israel Binational Science Foundation (BSF), by the Minerva foundation with funding from the Federal German Ministry for Education and Research, by the German Research Foundation through a German-Israeli Project Cooperation (DIP) grant ``Holography and the Swampland'', and by a research grant from Martin Eisenstein. OA is the Samuel Sebba Professorial Chair of Pure and Applied Physics. 
OM is supported by the ERC-COG grant NP-QFT No. 864583
``Non-perturbative dynamics of quantum fields: from new deconfined
phases of matter to quantum black holes'', by the MUR-FARE2020 grant
No. R20E8NR3HX ``The Emergence of Quantum Gravity from Strong Coupling Dynamics''. OM is also partially supported by
the INFN ``Iniziativa Specifica GAST''.

\appendix

\section{Special functions}
\label{app:Special functions}
We will use the notations\footnote{Note that this convention is more common in the literature concerning modular functions and transformations. In some of the literature concerning elliptic functions one uses $q' = e^{\pi i \tau} = \sqrt{q}$, even though it is denoted by $q$ there.}
\begin{equation}
    q = e^{2i\pi\tau} \;, \qquad p = e^{2\pi i \sigma} \;, \qquad z = e^{2\pi i u}\;.
\end{equation}

\paragraph{q-Pochhammer symbol}
The q-Pochhamemer symbol is
\begin{equation}
    (z;q)_n \equiv \prod_{k=0}^{n-1}(1-zq^k) \;,\qquad (z;q)_\infty \equiv \prod_{k=0}^\infty (1-zq^k)\;, \qquad \text{for } |q|<1 \;.
\end{equation}
There are also a series expansion and a Plethystic representation for $(z;q)_\infty$,
\begin{equation}
    (z;q)_\infty = \sum_{n=0}^\infty \frac{(-1)^n q^{\frac{1}{2}n(n-1)}}{(q;q)_n}z^n = \exp\left[-\sum_{k=1}^\infty \frac{1}{k} \frac{z^k}{1-q^k} \right] \;,
\end{equation}
where the first converges for $|q|<1$ while the second converges for $|z|,|q|<1$.

By relating the symbol to the Dedekind eta function,
\begin{equation}
    \eta(\tau) = e^{\frac{\pi i \tau}{12}}(q;q)_\infty \;,
\end{equation}
one obtains the properties of the q-Pochhammer symbol under modular transformations
\begin{equation}
    (\tilde q; \tilde q)_\infty = \sqrt{-i\tau} e^{\frac{\pi i}{12}(\tau + 1/\tau)}(q;q)_\infty \;,
\end{equation}
where $\tilde q = e^{-2\pi i/\tau}$. Finally, we have the asymptotic behaviour
\begin{equation}
    (z;q)_\infty \sim 1-z \qquad \text{for } q \to 0 \;, \qquad \qquad \log(z;q)_\infty \sim -\frac{z}{1-q} \quad \text{for } z\to 0 \;.
\end{equation}

\paragraph{Elliptic theta function}
The elliptic theta function is 
\begin{equation}
\label{eq:elliptic_theta_definition}
    \theta_0(u;\tau) \equiv (z;q)_\infty (q/z;q)_\infty = \prod_{k=0}^\infty (1-zq^k)(1-z^{-1}q^{k+1}) \; ,
\end{equation}
which gives an analytic function on $|q|<1$ with simple zeroes at $z=q^k$ for $k\in \bZ$ and no singularities. The infinite product is convergent in the whole domain. We can also give a plethystic expansion
\begin{equation}
\label{eq: plethystic definition theta}
    \theta_0(u;\tau) = \exp\left[ -\sum_{k=1}^\infty \frac{1}{k} \frac{z^k + (qz^{-1})^k}{1-q^k} \right] \;, 
\end{equation}
which converges for $|q| < |z| < 1$. The periodicity relations are 
\begin{equation}
\label{eq:elliptic theta periodicity}
\begin{aligned}
    \theta_0(u + n + m\tau; \tau) &= (-1)^m e^{-\pi i m (2  u + (m-1)\tau)}\theta_0(u;\tau) \;, \qquad\qquad m,n\in \bZ \\
    \theta_0(u;\tau) &= \theta_0(\tau - u;\tau) = -e^{2\pi i u} \theta_0(-u;\tau) \;,
\end{aligned}
\end{equation}
and under modular transformations
\begin{equation}
    \theta_0(u;\tau + 1) = \theta_0(u;\tau) \;, \qquad \theta_0\left(\frac{u}{\tau}; -\frac{1}{\tau} \right) = -ie^{\frac{\pi i}{\tau}(u^2 + u + \frac{1}{6}) - \pi i u + \frac{\pi i \tau}{6}}\theta_0(u;\tau) \;. 
\end{equation}

\paragraph{The elliptic Gamma function}
The elliptic Gamma function is defined by
\begin{equation}
    \wt \Gamma(u;\sigma,\tau) \equiv \prod_{m,n = 0}^\infty \frac{1-p^{m+1}q^{n+1}z^{-1}}{1-p^m q^n z} \; .
\end{equation}
This definition gives a meromorphic single valued function on $|p|,|q|<1$ with simple zeroes at $z = p^{m+1}q^{n+1}$ and simple poles at $z = p^{-m}q^{-n}$ for $m,n \ge 0$. The infinite product is convergent on the whole domain. We can also give a plethystic definition
\begin{equation}
    \wt \Gamma(u;\sigma,\tau) = \exp\left[\sum_{k=1}^\infty \frac{1}{k}\frac{z^k - (pqz^{-1})^k}{(1-p^k)(1-q^k)} \right] \;,
\end{equation}
which converges for $|pq|<|z|<1$. The function has the following periodicity relations:
\begin{equation}
\label{eq:elliptic gamma quasi periodicity}
\begin{aligned}
    \wt\Gamma(u;\sigma,\tau) &= \wt\Gamma(u;\tau,\sigma) \;, \\
    \wt\Gamma(u;\sigma,\tau) = \wt\Gamma(u+1;\sigma,\tau) &= \wt\Gamma(u;\sigma+1,\tau) = \wt\Gamma(u;\sigma,\tau+1) \;, \\
    \wt\Gamma(u+\sigma; \sigma, \tau) &= \theta_0(u;\tau) \wt\Gamma(u;\sigma,\tau) \;,  \\
    \qquad \wt\Gamma(u+\tau;\sigma,\tau) &= \theta_0(u;\sigma) \wt\Gamma(u;\sigma,\tau) \;.
\end{aligned}
\end{equation}

Moreover,
\begin{equation}
\label{eq: inversion formula}
    \wt\Gamma(u;\sigma,\tau) \wt\Gamma(\sigma + \tau - u;\sigma,\tau) = 1 \;.
\end{equation}
The elliptic Gamma function has $SL(3,\bZ)$ modular properties. For $\sigma, \tau, \sigma/\tau, \sigma+\tau \in \bC \setminus \cR$ there is a "modular formula": \cite{Felder:2000}
\begin{equation}
    \wt\Gamma(u;\sigma,\tau) = e^{-\pi i \cQ(u;\sigma,\tau)}\frac{\wt\Gamma\left(\frac{u}{\tau}; \frac{\sigma}{\tau}, -\frac{1}{\tau}\right)}{\wt\Gamma\left(\frac{u-\tau}{\sigma}; -\frac{1}{\sigma}, -\frac{\tau}{\sigma}\right)} = e^{-\pi i \cQ(u;\sigma,\tau)}\frac{\wt\Gamma\left(\frac{u}{\sigma}; -\frac{1}{\sigma}, \frac{\tau}{\sigma}\right)}{\wt\Gamma\left(\frac{u-\sigma}{\tau}; -\frac{\sigma}{\tau}, -\frac{1}{\tau}\right)} \;,
\end{equation}
where $\cQ(u;\sigma,\tau)$ is the cubic polynomial
\begin{equation}
    \cQ(u;\sigma,\tau) = \frac{u^3}{3\sigma\tau} -\frac{\sigma+\tau-1}{2\sigma\tau}u^2 +
    \frac{\sigma^2+\tau^2+3\sigma\tau-3\sigma-3\tau+1}{6\sigma\tau}u + \frac{(\sigma+\tau-1)(\sigma+\tau-\sigma\tau)}{12\sigma\tau} \;.
\end{equation}

In the degenerate case $\sigma = \tau$ the formula above is not valid. For $u \in \bC \setminus (\bZ + \tau \bZ)$, however, there is a degenerate relation
\begin{equation}
\label{eq: elliptic gamma function decomposition}
    \wt\Gamma(u;\tau,\tau) = \frac{e^{-\pi i \cQ(u;\tau,\tau)}}{\theta_0(\frac{u}{\tau};-\frac{1}{\tau})}\prod_{k=0}^\infty\frac{\psi\left(\frac{k+1+u}{\tau}\right)}{\psi\left(\frac{k-u}{\tau}\right)} \;,
\end{equation}
where the function $\psi$ is the elliptic digamma function defined below and the polynomial $\cQ$ reduces to 
\begin{equation}
\label{eq:def Q modular}
    \cQ(u;\tau,\tau) = \frac{(2u-2\tau+1)(2u(u+1)-2\tau(2u+1)+\tau^2)}{12\tau^2}\;.
\end{equation}
Using 
\begin{equation}
    \cQ(u+1;\tau,\tau) - \cQ(u;\tau,\tau) = \frac{(u+1)(u+1-2\tau)}{\tau^2}+\frac{5}{6} \;,
\end{equation}
one can check that $\wt\Gamma(u;\tau,\tau)$ is invariant under $u\to u+1$.

%

\paragraph{Function \matht{\psi}.} Define, for $\im (t) < 0$, the function
\be
\label{function psi definition}
\psi(t) = \exp \left[ t \log \bigl( 1-e^{-2\pi i t} \bigr) - \frac1{2\pi i} \text{Li}_2 (e^{-2\pi i t}) \right] 
= \exp \left[ - \sum_{\ell=1}^\infty \left( \frac t\ell + \frac1{2\pi i \, \ell^2} \right) e^{-2\pi i t \, \ell} \right] \;.
\ee
The branch of the logarithm is determined by its series expansion $\log(1-z) = - \sum_{\ell=1}^\infty z^\ell/ \ell$, whereas $\text{Li}_2(z) = \sum_{\ell=1}^\infty z^\ell / \ell^2$ is the dilogarithm. One can show that the branch cut discontinuities of the logarithm and the dilogarithm cancel in the definition of $\psi(t)$, such that the latter extends to a meromorphic function on the whole complex plane. Some useful properties of $\psi(t)$ are:
\be
\psi(t) \, \psi(-t) = e^{-\pi i (t^2 - 1/6)} \;,\qquad\qquad \psi(t+n) = (1-e^{-2\pi i t})^n \, \psi(t) \qquad\text{for}\quad n \in \bZ \;.
\ee
In particular, from (\ref{function psi definition}), $\psi(0) = e^{\pi i / 12}$.

\printbibliography

@article{Felder:2000,
   title={The Elliptic Gamma Function and SL(3,$\mathbb{Z}$)$\ltimes Z_3$},
   volume={156},
   ISSN={0001-8708},
   url={http://dx.doi.org/10.1006/aima.2000.1951},
   DOI={10.1006/aima.2000.1951},
  eprint		= "math/9907061",
  archivePrefix = "arXiv",
  primaryClass = "math.QA",
   number={1},
   journal={Advances in Mathematics},
   publisher={Elsevier BV},
   author={Felder, Giovanni and Varchenko, Alexander},
   year={2000},
   pages={44-76}
}

@article{Aharony:2021zkr,
    author = "Aharony, Ofer and Benini, Francesco and Mamroud, Ohad and Milan, Elisa",
    title = "{A gravity interpretation for the Bethe Ansatz expansion of the $\mathcal{N}=4$ SYM index}",
    eprint = "2104.13932",
    archivePrefix = "arXiv",
    primaryClass = "hep-th",
    reportNumber = "SISSA 01/2021/FISI",
    doi = "10.1103/PhysRevD.104.086026",
    journal = "Phys. Rev. D",
    volume = "104",
    pages = "086026",
    year = "2021"
}

@article{ArabiArdehali:2021nsx,
    author = "Arabi Ardehali, Arash and Murthy, Sameer",
    title = "{The 4d superconformal index near roots of unity and 3d Chern-Simons theory}",
    eprint = "2104.02051",
    archivePrefix = "arXiv",
    primaryClass = "hep-th",
    doi = "10.1007/JHEP10(2021)207",
    journal = "JHEP",
    volume = "10",
    pages = "207",
    year = "2021"
}

@article{GonzalezLezcano:2020yeb,
    author = "Gonz\'alez Lezcano, Alfredo and Hong, Junho and Liu, James T. and Pando Zayas, Leopoldo A.",
    title = "{Sub-leading Structures in Superconformal Indices: Subdominant Saddles and Logarithmic Contributions}",
    eprint = "2007.12604",
    archivePrefix = "arXiv",
    primaryClass = "hep-th",
    reportNumber = "LCTP-20-16",
    doi = "10.1007/JHEP01(2021)001",
    journal = "JHEP",
    volume = "01",
    pages = "001",
    year = "2021"
}

@article{Kim:2019yrz,
    author = "Kim, Joonho and Kim, Seok and Song, Jaewon",
    title = "{A 4d $ \mathcal{N} $ = 1 Cardy Formula}",
    eprint = "1904.03455",
    archivePrefix = "arXiv",
    primaryClass = "hep-th",
    reportNumber = "KIAS-P19015, SNUTP19-002",
    doi = "10.1007/JHEP01(2021)025",
    journal = "JHEP",
    volume = "01",
    pages = "025",
    year = "2021"
}

@article{Cabo-Bizet:2019osg,
    author = "Cabo-Bizet, Alejandro and Cassani, Davide and Martelli, Dario and Murthy, Sameer",
    title = "{The asymptotic growth of states of the 4d $ \mathcal{N}=1 $ superconformal index}",
    eprint = "1904.05865",
    archivePrefix = "arXiv",
    primaryClass = "hep-th",
    doi = "10.1007/JHEP08(2019)120",
    journal = "JHEP",
    volume = "08",
    pages = "120",
    year = "2019"
}

@article{Benini:2018ywd,
    author = "Benini, Francesco and Milan, Elisa",
    title = "{Black Holes in 4D $\mathcal{N}$=4 Super-Yang-Mills Field Theory}",
    eprint = "1812.09613",
    archivePrefix = "arXiv",
    primaryClass = "hep-th",
    reportNumber = "SISSA 56/2018/FISI",
    doi = "10.1103/PhysRevX.10.021037",
    journal = "Phys. Rev. X",
    volume = "10",
    number = "2",
    pages = "021037",
    year = "2020"
}

@article{Cabo-Bizet:2018ehj,
    author = "Cabo-Bizet, Alejandro and Cassani, Davide and Martelli, Dario and Murthy, Sameer",
    title = "{Microscopic origin of the Bekenstein-Hawking entropy of supersymmetric AdS$_{5}$ black holes}",
    eprint = "1810.11442",
    archivePrefix = "arXiv",
    primaryClass = "hep-th",
    doi = "10.1007/JHEP10(2019)062",
    journal = "JHEP",
    volume = "10",
    pages = "062",
    year = "2019"
}

@article{Choi:2018hmj,
    author = "Choi, Sunjin and Kim, Joonho and Kim, Seok and Nahmgoong, June",
    title = "{Large AdS black holes from QFT}",
    eprint = "1810.12067",
    archivePrefix = "arXiv",
    primaryClass = "hep-th",
    reportNumber = "SNUTP18-005, KIAS-P18097",
    month = "10",
    year = "2018"
}

@article{Hong:2018viz,
    author = "Hong, Junho and Liu, James T.",
    title = "{The topologically twisted index of $ \mathcal{N} $ = 4 super-Yang-Mills on T$^{2} \times S^{2}$ and the elliptic genus}",
    eprint = "1804.04592",
    archivePrefix = "arXiv",
    primaryClass = "hep-th",
    reportNumber = "LCTP-18-12",
    doi = "10.1007/JHEP07(2018)018",
    journal = "JHEP",
    volume = "07",
    pages = "018",
    year = "2018"
}

@article{Cabo-Bizet:2019eaf,
    author = "Cabo-Bizet, Alejandro and Murthy, Sameer",
    title = "{Supersymmetric phases of 4d $ \mathcal{N} $ = 4 SYM at large $N$}",
    eprint = "1909.09597",
    archivePrefix = "arXiv",
    primaryClass = "hep-th",
    doi = "10.1007/JHEP09(2020)184",
    journal = "JHEP",
    volume = "09",
    pages = "184",
    year = "2020"
}

@article{Benini:2018mlo,
    author = "Benini, Francesco and Milan, Elisa",
    title = "{A Bethe Ansatz type formula for the superconformal index}",
    eprint = "1811.04107",
    archivePrefix = "arXiv",
    primaryClass = "hep-th",
    reportNumber = "SISSA 46/2018/FISI",
    doi = "10.1007/s00220-019-03679-y",
    journal = "Commun. Math. Phys.",
    volume = "376",
    number = "2",
    pages = "1413--1440",
    year = "2020"
}

@misc{Aharony:2024,
    author = "Aharony, Ofer and Benini, Francesco and Mamroud, Ohad",
    howpublished = "work in progress",
}

@article{Choi:2021rxi,
    author = "Choi, Sunjin and Jeong, Saebyeok and Kim, Seok and Lee, Eunwoo",
    title = "{Exact QFT duals of AdS black holes}",
    eprint = "2111.10720",
    archivePrefix = "arXiv",
    primaryClass = "hep-th",
    reportNumber = "KIAS-P21054, SNUTP21-002",
    month = "11",
    year = "2021"
}

@article{Agarwal:2020zwm,
      author = "Agarwal, Prarit and Choi, Sunjin and Kim, Joonho and Kim, Seok and Nahmgoong, June",
      title = "{AdS black holes and finite $N$ indices}",
      eprint = "2005.11240",
      archivePrefix = "arXiv",
      primaryClass = "hep-th",
      year = "2020",
}

@article{Amariti:2019mgp,
      author         = "Amariti, Antonio and Garozzo, Ivan and Lo Monaco, Gabriele",
      title          = "{Entropy function from toric geometry}",
      year           = "2019",
      eprint         = "1904.10009",
      archivePrefix  = "arXiv",
      primaryClass   = "hep-th",
      SLACcitation   = "%%CITATION = ARXIV:1904.10009;%%"
}

@article{Amariti:2020jyx,
      author = "Amariti, Antonio and Fazzi, Marco and Segati, Alessia",
      title = "{The SCI of $\mathcal{N}{=}4$ $USp(2N_c)$ and $SO(N_c)$ SYM as a matrix integral}",
      eprint = "2012.15208",
      archivePrefix = "arXiv",
      primaryClass = "hep-th",
      month = "12",
      year = "2020",
}

@article{ArabiArdehali:2019tdm,
      author = "Arabi Ardehali, Arash",
      title = "{Cardy-like asymptotics of the 4d $\mathcal{N}{=}4$ index and AdS$_5$ blackholes}",
      journal = "JHEP",
      volume = "06",
      pages = "134",
      year = "2019",
      doi = "10.1007/JHEP06(2019)134",
      eprint = "1902.06619",
      archivePrefix = "arXiv",
      primaryClass = "hep-th",
}

@article{ArabiArdehali:2019orz,
      author         = "Arabi Ardehali, Arash and Hong, Junho and Liu, James T.",
      title          = "{Asymptotic growth of the 4d $\mathcal{N}{=}4$ index and
                        partially deconfined phases}",
      journal = "JHEP",
      volume = "07",
      pages = "073",
      year = "2020",
      doi = "10.1007/JHEP07(2020)073",
      eprint         = "1912.04169",
      archivePrefix  = "arXiv",
      primaryClass   = "hep-th",
      SLACcitation   = "%%CITATION = ARXIV:1912.04169;%%"
}

@article{Azzurli:2017kxo,
      author = "Azzurli, Francesco and Bobev, Nikolay and Crichigno, P. Marcos and Min, Vincent S. and Zaffaroni, Alberto",
      title = "{A universal counting of black hole microstates in AdS$_{4}$}",
      journal = "JHEP",
      volume = "02",
      pages = "054",
      year = "2018",
      doi = "10.1007/JHEP02(2018)054",
      eprint = "1707.04257",
      archivePrefix = "arXiv",
      primaryClass = "hep-th",
}

@article{Benini:2015eyy,
      author         = "Benini, Francesco and Hristov, Kiril and Zaffaroni, Alberto",
      title          = "{Black hole microstates in AdS$_{4}$ from supersymmetric localization}",
      journal        = "JHEP",
      volume         = "05",
      year           = "2016",
      pages          = "054",
      doi            = "10.1007/JHEP05(2016)054",
      eprint         = "1511.04085",
      archivePrefix  = "arXiv",
      primaryClass   = "hep-th",
      SLACcitation   = "%%CITATION = ARXIV:1511.04085;%%"
}

@article{Benini:2016hjo,
      author		= "Benini, Francesco and Zaffaroni, Alberto",
      title		= "{Supersymmetric partition functions on Riemann surfaces}",
      journal		= "Proc. Symp. Pure Math.",
      volume	= "96",
      pages		= "13-46",
      year		= "2017",
      doi		= "10.1090/pspum/096",
      editor		= "Li, Si and Lian, Bong H. and Song, Wei and Yau, Shing-Tung",
      eprint		= "1605.06120",
      archivePrefix = "arXiv",
      primaryClass = "hep-th",
      SLACcitation   = "%%CITATION = ARXIV:1605.06120;%%"
}

@article{Benini:2016rke,
      author         = "Benini, Francesco and Hristov, Kiril and Zaffaroni, Alberto",
      title          = "{Exact microstate counting for dyonic black holes in AdS$_4$}",
      journal        = "Phys. Lett.",
      volume         = "B771",
      year           = "2017",
      pages          = "462-466",
      doi            = "10.1016/j.physletb.2017.05.076",
      eprint         = "1608.07294",
      archivePrefix  = "arXiv",
      primaryClass   = "hep-th",
      SLACcitation   = "%%CITATION = ARXIV:1608.07294;%%"
}

@article{Benini:2017oxt,
      author         = "Benini, Francesco and Khachatryan, Hrachya and Milan, Paolo",
      title          = "{Black hole entropy in massive Type IIA}",
      journal        = "Class. Quant. Grav.",
      volume         = "35",
      year           = "2018",
      pages          = "035004",
      doi            = "10.1088/1361-6382/aa9f5b",
      eprint         = "1707.06886",
      archivePrefix  = "arXiv",
      primaryClass   = "hep-th",
      SLACcitation   = "%%CITATION = ARXIV:1707.06886;%%"
}

@article{Benini:2019dyp,
      author = "Benini, Francesco and Gang, Dongmin and Pando Zayas, Leopoldo A.",
      title = "{Rotating Black Hole Entropy from M5-branes}",
      journal = "JHEP",
      volume = "03",
      pages = "057",
      year = "2020",
      doi = "10.1007/JHEP03(2020)057",
      eprint = "1909.11612",
      archivePrefix = "arXiv",
      primaryClass = "hep-th",
}

@article{Benini:2020gjh,
      author = "Benini, Francesco and Colombo, Edoardo and Soltani, Saman and Zaffaroni, Alberto and Zhang, Ziruo",
      title = "{Superconformal indices at large $N$ and the entropy of AdS$_5$ $\times$ SE$_5$ black holes}",
      journal = "Class. Quant. Grav.",
      volume = "37",
      number = "21",
      pages = "215021",
      year = "2020",
      doi = "10.1088/1361-6382/abb39b",
      eprint = "2005.12308",
      archivePrefix = "arXiv",
      primaryClass = "hep-th",
}

@article{Benini:2021ano,
      author = "Benini, Francesco and Rizi, Giovanni",
      title = "{Superconformal index of low-rank gauge theories via the Bethe Ansatz}",
      journal = "JHEP",
      volume = "05",
      pages = "061",
      year = "2021",
      doi = "10.1007/JHEP05(2021)061",
      eprint = "2102.03638",
      archivePrefix = "arXiv",
      primaryClass = "hep-th",
}

@article{Bobev:2019zmz,
      author = "Bobev, Nikolay and Crichigno, P. Marcos",
      title = "{Universal spinning black holes and theories of class~$\mathcal{R}$}",
      journal = "JHEP",
      volume = "12",
      pages = "054",
      year = "2019",
      doi = "10.1007/JHEP12(2019)054",
      eprint = "1909.05873",
      archivePrefix = "arXiv",
      primaryClass = "hep-th",
}

@article{Cabo-Bizet:2017jsl,
      author = "Cabo-Bizet, Alejandro and Giraldo-Rivera, Victor I. and Pando Zayas, Leopoldo A.",
      title = "{Microstate counting of AdS$_{4}$ hyperbolic black hole entropy via the topologically twisted index}",
      journal = "JHEP",
      volume = "08",
      pages = "023",
      year = "2017",
      doi = "10.1007/JHEP08(2017)023",
      eprint = "1701.07893",
      archivePrefix = "arXiv",
      primaryClass = "hep-th",
}

@article{Cabo-Bizet:2020nkr,
      author = "Cabo-Bizet, Alejandro and Cassani, Davide and Martelli, Dario and Murthy, Sameer",
      title = "{The large-$N$ limit of the 4d $\mathcal{N}{=}1$ superconformal index}",
      journal = "JHEP",
      volume = "11",
      pages = "150",
      year = "2020",
      doi = "10.1007/JHEP11(2020)150",
      eprint = "2005.10654",
      archivePrefix = "arXiv",
      primaryClass = "hep-th",
}

@article{Cabo-Bizet:2020ewf,
      author = "Cabo-Bizet, Alejandro",
      title = "{From multi-gravitons to Black holes: The role of complex saddles}",
      eprint = "2012.04815",
      archivePrefix = "arXiv",
      primaryClass = "hep-th",
      month = "12",
      year = "2020",
}

@article{Choi:2019zpz,
      author = "Choi, Sunjin and Hwang, Chiung and Kim, Seok",
      title = "{Quantum vortices, M2-branes and black holes}",
      year = "2019",
      eprint = "1908.02470",
      archivePrefix = "arXiv",
      primaryClass = "hep-th",
}

@article{Closset:2017bse,
      author         = "Closset, Cyril and Kim, Heeyeon and Willett, Brian",
      title          = "{$\mathcal{N}{=}1$ supersymmetric indices and the four-dimensional $A$-model}",
      journal        = "JHEP",
      volume         = "08",
      year           = "2017",
      pages          = "090",
      doi            = "10.1007/JHEP08(2017)090",
      eprint         = "1707.05774",
      archivePrefix  = "arXiv",
      primaryClass   = "hep-th",
      SLACcitation   = "%%CITATION = ARXIV:1707.05774;%%"
}

@article{Copetti:2020dil,
      author = "Copetti, Christian and Grassi, Alba and Komargodski, Zohar and Tizzano, Luigi",
      title = "{Delayed Deconfinement and the Hawking-Page Transition}",
      year = "2020",
      eprint = "2008.04950",
      archivePrefix = "arXiv",
      primaryClass = "hep-th",
}

@article{Crichigno:2018adf,
      author = "Crichigno, P. Marcos and Jain, Dharmesh and Willett, Brian",
      title = "{5d Partition Functions with A Twist}",
      journal = "JHEP",
      volume = "11",
      pages = "058",
      year = "2018",
      doi = "10.1007/JHEP11(2018)058",
      eprint = "1808.06744",
      archivePrefix = "arXiv",
      primaryClass = "hep-th",
}

@article{Fluder:2019szh,
      author = "Fluder, Martin and Hosseini, Seyed Morteza and Uhlemann, Christoph F.",
      title = "{Black hole microstate counting in Type IIB from 5d SCFTs}",
      journal = "JHEP",
      volume = "05",
      pages = "134",
      year = "2019",
      doi = "10.1007/JHEP05(2019)134",
      eprint = "1902.05074",
      archivePrefix = "arXiv",
      primaryClass = "hep-th",
}

@article{Gang:2019uay,
      author = "Gang, Dongmin and Kim, Nakwoo and Pando Zayas, Leopoldo A.",
      title = "{Precision Microstate Counting for the Entropy of Wrapped M5-branes}",
      journal = "JHEP",
      volume = "03",
      pages = "164",
      year = "2020",
      doi = "10.1007/JHEP03(2020)164",
      eprint = "1905.01559",
      archivePrefix = "arXiv",
      primaryClass = "hep-th",
}

@article{Goldstein:2020yvj,
      author = "Goldstein, Kevin and Jejjala, Vishnu and Lei, Yang and van Leuven, Sam and Li, Wei",
      title = "{Residues, modularity, and the Cardy limit of the 4d $\mathcal{N}{=}4$ superconformal index}",
      journal = "JHEP",
      volume = "04",
      pages = "216",
      year = "2021",
      doi = "10.1007/JHEP04(2021)216",
      eprint = "2011.06605",
      archivePrefix = "arXiv",
      primaryClass = "hep-th",
}

@article{Honda:2019cio,
      author = "Honda, Masazumi",
      title = "{Quantum Black Hole Entropy from 4d Supersymmetric Cardy formula}",
      journal = "Phys. Rev. D",
      volume = "100",
      number = "2",
      pages = "026008",
      year = "2019",
      doi = "10.1103/PhysRevD.100.026008",
      eprint = "1901.08091",
      archivePrefix = "arXiv",
      primaryClass = "hep-th",
}

@article{Hosseini:2016tor,
      author = "Hosseini, Seyed Morteza and Zaffaroni, Alberto",
      title = "{Large $N$ matrix models for 3d $\mathcal{N}{=}2$ theories: twisted index, free energy and black holes}",
      journal = "JHEP",
      volume = "08",
      pages = "064",
      year = "2016",
      doi = "10.1007/JHEP08(2016)064",
      eprint = "1604.03122",
      archivePrefix = "arXiv",
      primaryClass = "hep-th",
}

@article{Hosseini:2016cyf,
      author         = "Hosseini, Seyed Morteza and Nedelin, Anton and Zaffaroni, Alberto",
      title          = "{The Cardy limit of the topologically twisted index and black strings in AdS$_{5}$}",
      journal        = "JHEP",
      volume         = "04",
      year           = "2017",
      pages          = "014",
      doi            = "10.1007/JHEP04(2017)014",
      eprint         = "1611.09374",
      archivePrefix  = "arXiv",
      primaryClass   = "hep-th",
      SLACcitation   = "%%CITATION = ARXIV:1611.09374;%%"
}

@article{Hosseini:2017fjo,
      author         = "Hosseini, Seyed Morteza and Hristov, Kiril and Passias, Achilleas",
      title          = "{Holographic microstate counting for AdS$_{4}$ black holes in massive IIA supergravity}",
      journal        = "JHEP",
      volume         = "10",
      year           = "2017",
      pages          = "190",
      doi            = "10.1007/JHEP10(2017)190",
      eprint         = "1707.06884",
      archivePrefix  = "arXiv",
      primaryClass   = "hep-th",
      SLACcitation   = "%%CITATION = ARXIV:1707.06884;%%"
}

@article{Hosseini:2018uzp,
      author = "Hosseini, Seyed Morteza and Yaakov, Itamar and Zaffaroni, Alberto",
      title = "{Topologically twisted indices in five dimensions and holography}",
      journal = "JHEP",
      volume = "11",
      pages = "119",
      year = "2018",
      doi = "10.1007/JHEP11(2018)119",
      eprint = "1808.06626",
      archivePrefix = "arXiv",
      primaryClass = "hep-th",
}

@article{Hosseini:2018usu,
      author = "Hosseini, Seyed Morteza and Hristov, Kiril and Passias, Achilleas and Zaffaroni, Alberto",
      title = "{6D attractors and black hole microstates}",
      journal = "JHEP",
      volume = "12",
      pages = "001",
      year = "2018",
      doi = "10.1007/JHEP12(2018)001",
      eprint = "1809.10685",
      archivePrefix = "arXiv",
      primaryClass = "hep-th",
}

@article{Jejjala:2021hlt,
      author = "Jejjala, Vishnu and Lei, Yang and Van Leuven, Sam and Li, Wei",
      title = "{$SL(3,\mathbb{Z})$ Modularity and New Cardy Limits of the $\mathcal{N}{=}4$ Superconformal Index}",
      year = "2021",
      eprint = "2104.07030",
      archivePrefix = "arXiv",
      primaryClass = "hep-th",
}

@article{Kantor:2019lfo,
      author = "K\'antor, Gergely and Papageorgakis, Constantinos and Richmond, Paul",
      title = "{AdS$_{7}$ black-hole entropy and 5D $\mathcal{N}{=}2$ Yang-Mills}",
      journal = "JHEP",
      volume = "01",
      pages = "017",
      year = "2020",
      doi = "10.1007/JHEP01(2020)017",
      eprint = "1907.02923",
      archivePrefix = "arXiv",
      primaryClass = "hep-th",
}

@article{Kinney:2005ej,
      author         = "Kinney, Justin and Maldacena, Juan Martin and Minwalla,
                        Shiraz and Raju, Suvrat",
      title          = "{An Index for 4 dimensional super conformal theories}",
      journal        = "Commun. Math. Phys.",
      volume         = "275",
      year           = "2007",
      pages          = "209-254",
      doi            = "10.1007/s00220-007-0258-7",
      eprint         = "hep-th/0510251",
      archivePrefix  = "arXiv",
      primaryClass   = "hep-th",
      SLACcitation   = "%%CITATION = HEP-TH/0510251;%%"
}

@article{Lanir:2019abx,
      author		= "Lanir, Assaf and Nedelin, Anton and Sela, Orr",
      title		= "{Black hole entropy function for toric theories via Bethe Ansatz}",
      journal		= "JHEP",
      volume	= "04",
      pages		= "091",
      year		= "2020",
      doi		= "10.1007/JHEP04(2020)091",
      eprint         = "1908.01737",
      archivePrefix  = "arXiv",
      primaryClass   = "hep-th",
      SLACcitation   = "%%CITATION = ARXIV:1908.01737;%%"
}

@article{Lezcano:2019pae,
      author = "Gonz\'alez Lezcano, Alfredo and Pando Zayas, Leopoldo A.",
      title = "{Microstate counting via Bethe Ans\"{a}tze in the 4d $\mathcal{N}{=}1$ superconformal index}",
      journal = "JHEP",
      volume = "03",
      pages = "088",
      year = "2020",
      doi = "10.1007/JHEP03(2020)088",
      eprint = "1907.12841v3",
      archivePrefix = "arXiv",
      primaryClass = "hep-th",
}

@article{Murthy:2020rbd,
      author = "Murthy, Sameer",
      title = "{The growth of the $\frac{1}{16}$-BPS index in 4d $\mathcal{N}{=}4$ SYM}",
      eprint = "2005.10843",
      archivePrefix = "arXiv",
      primaryClass = "hep-th",
      year = "2020",
}

@article{Nian:2019pxj,
      author = "Nian, Jun and Pando Zayas, Leopoldo A.",
      title = "{Microscopic entropy of rotating electrically charged AdS$_{4}$ black holes from field theory localization}",
      journal = "JHEP",
      volume = "03",
      pages = "081",
      year = "2020",
      doi = "10.1007/JHEP03(2020)081",
      eprint = "1909.07943",
      archivePrefix = "arXiv",
      primaryClass = "hep-th",
}

@article{Romelsberger:2005eg,
      author         = "Romelsberger, Christian",
      title          = "{Counting chiral primaries in $\mathcal{N}{=}1$, $d{=}4$ superconformal field theories}",
      journal        = "Nucl. Phys.",
      volume         = "B747",
      year           = "2006",
      pages          = "329-353",
      doi            = "10.1016/j.nuclphysb.2006.03.037",
      eprint         = "hep-th/0510060",
      archivePrefix  = "arXiv",
      primaryClass   = "hep-th",
      SLACcitation   = "%%CITATION = HEP-TH/0510060;%%"
}

@article{Suh:2018szn,
      author = "Suh, Minwoo",
      title = "{Supersymmetric AdS$_6$ black holes from matter coupled $F(4)$ gauged supergravity}",
      journal = "JHEP",
      volume = "02",
      pages = "108",
      year = "2019",
      eprint = "1810.00675",
      archivePrefix = "arXiv",
      primaryClass = "hep-th",
      doi = "10.1007/JHEP02(2019)108",
}

@article{David:2021qaa,
    author = "David, Marina and Lezcano Gonz\'alez, Alfredo and Nian, Jun and Pando Zayas, Leopoldo A.",
    title = "{Logarithmic corrections to the entropy of rotating black holes and black strings in AdS$_{5}$}",
    eprint = "2106.09730",
    archivePrefix = "arXiv",
    primaryClass = "hep-th",
    reportNumber = "LCTP-21-14",
    doi = "10.1007/JHEP04(2022)160",
    journal = "JHEP",
    volume = "04",
    pages = "160",
    year = "2022"
}

@article{David:2020ems,
    author = "David, Marina and Nian, Jun and Pando Zayas, Leopoldo A.",
    title = "{Gravitational Cardy Limit and AdS Black Hole Entropy}",
    eprint = "2005.10251",
    archivePrefix = "arXiv",
    primaryClass = "hep-th",
    reportNumber = "LCTP-20-09",
    doi = "10.1007/JHEP11(2020)041",
    journal = "JHEP",
    volume = "11",
    pages = "041",
    year = "2020"
}

@article{Amariti:2022nvn,
    author = "Amariti, Antonio and Segati, Alessia",
    title = "{Kerr-Newman black holes from $\mathcal{N}=1^*$}",
    eprint = "2210.03015",
    archivePrefix = "arXiv",
    primaryClass = "hep-th",
    month = "10",
    year = "2022"
}

@article{Hosseini:2021mnn,
    author = "Hosseini, Seyed Morteza and Yaakov, Itamar and Zaffaroni, Alberto",
    title = "{The joy of factorization at large N: five-dimensional indices and AdS black holes}",
    eprint = "2111.03069",
    archivePrefix = "arXiv",
    primaryClass = "hep-th",
    doi = "10.1007/JHEP02(2022)097",
    journal = "JHEP",
    volume = "02",
    pages = "097",
    year = "2022"
}

@article{Colombo:2021kbb,
    author = "Colombo, Edoardo",
    title = "{The large-N limit of 4d superconformal indices for general BPS charges}",
    eprint = "2110.01911",
    archivePrefix = "arXiv",
    primaryClass = "hep-th",
    doi = "10.1007/JHEP12(2022)013",
    journal = "JHEP",
    volume = "12",
    pages = "013",
    year = "2022"
}

@article{Hong:2021bzg,
    author = "Hong, Junho",
    title = "{The topologically twisted index of $ \mathcal{N} $ = 4 SU(N) Super-Yang-Mills theory and a black hole Farey tail}",
    eprint = "2108.02355",
    archivePrefix = "arXiv",
    primaryClass = "hep-th",
    reportNumber = "LCTP-21-20",
    doi = "10.1007/JHEP10(2021)145",
    journal = "JHEP",
    volume = "10",
    pages = "145",
    year = "2021"
}

@article{Jejjala:2022lrm,
    author = "Jejjala, Vishnu and Lei, Yang and van Leuven, Sam and Li, Wei",
    title = "{Modular factorization of superconformal indices}",
    eprint = "2210.17551",
    archivePrefix = "arXiv",
    primaryClass = "hep-th",
    month = "10",
    year = "2022"
}

@article{Mamroud:2022msu,
    author = "Mamroud, Ohad",
    title = "{The SUSY Index Beyond the Cardy Limit}",
    eprint = "2212.11925",
    archivePrefix = "arXiv",
    primaryClass = "hep-th",
    month = "12",
    year = "2022"
}

@article{BenettiGenolini:2023rkq,
    author = "Benetti Genolini, Pietro and Cabo-Bizet, Alejandro and Murthy, Sameer",
    title = "{Supersymmetric phases of AdS$_{4}$/CFT$_{3}$}",
    eprint = "2301.00763",
    archivePrefix = "arXiv",
    primaryClass = "hep-th",
    doi = "10.1007/JHEP06(2023)125",
    journal = "JHEP",
    volume = "06",
    pages = "125",
    year = "2023"
}

@article{Choi:2023tiq,
    author = "Choi, Sunjin and Kim, Seungkyu and Song, Jaewon",
    title = "{Large $N$ Universality of 4d $\mathcal{N}=1$ Superconformal Index and AdS Black Holes}",
    eprint = "2309.07614",
    archivePrefix = "arXiv",
    primaryClass = "hep-th",
    reportNumber = "KIAS-P23034",
    month = "9",
    year = "2023"
}

@article{Bobev:2023dwx,
    author = "Bobev, Nikolay and David, Marina and Hong, Junho and Reys, Valentin and Zhang, Xuao",
    title = "{A compendium of logarithmic corrections in AdS/CFT}",
    eprint = "2312.08909",
    archivePrefix = "arXiv",
    primaryClass = "hep-th",
    month = "12",
    year = "2023"
}

@article{Cabo-Bizet:2023ejm,
    author = "Cabo-Bizet, Alejandro and David, Marina and Gonz\'alez Lezcano, Alfredo",
    title = "{Thermodynamics of black holes with probe D-branes}",
    eprint = "2312.12533",
    archivePrefix = "arXiv",
    primaryClass = "hep-th",
    month = "12",
    year = "2023"
}

@article{Chen:2023lzq,
    author = "Chen, Yiming and Heydeman, Matthew and Wang, Yifan and Zhang, Mengyang",
    title = "{Probing supersymmetric black holes with surface defects}",
    eprint = "2306.05463",
    archivePrefix = "arXiv",
    primaryClass = "hep-th",
    reportNumber = "PUPT-2642",
    doi = "10.1007/JHEP10(2023)136",
    journal = "JHEP",
    volume = "10",
    pages = "136",
    year = "2023"
}
\end{document}